\documentclass[twocolumn]{aastex631}

\usepackage{enumitem}
\usepackage{xspace}
\usepackage{multirow}

\usepackage[encapsulated]{CJK}

\newcommand*{\JWST}{\emph{JWST}\xspace}

\newcommand{\msol}{M$_{\odot}$}
\newcommand{\zsol}{Z$_{\odot}$}

\newcommand{\pf}{\texttt{pyPlatefit}}

\newcommand{\lya}{Lyman-$\alpha$ }
\newcommand{\lyans}{Lyman-$\alpha$}

\newcommand{\nii}{[\ion{N}{2}]}

\newcommand{\hb}{H$\beta$}
\newcommand{\oiii}{[\ion{O}{3}]}
\newcommand{\oii}{[\ion{O}{2}]}

\newcommand{\ha}{H$\alpha$}

\shorttitle{JWST/NIRSpec Measurements of Extremely Low Z in high-EW LAEs}
\shortauthors{Maseda et al.}

\begin{document}

\title{JWST/NIRSpec Measurements of Extremely Low Metallicities in High Equivalent Width Lyman-$\alpha$ Emitters}

\author[0000-0003-0695-4414]{Michael V. Maseda}
\affiliation{Department of Astronomy, University of Wisconsin-Madison, 475 N. Charter St., Madison, WI 53706, USA}
\author{Zach Lewis}
\affiliation{Department of Astronomy, University of Wisconsin-Madison, 475 N. Charter St., Madison, WI 53706, USA}
\author[0000-0003-2871-127X]{Jorryt Matthee}
\affil{ETH Z\"urich, Department of Physics, Wolfgang-Pauli-Str. 27, 8093 Z\"urich, Switzerland}
\author[0000-0002-7054-4332]{Joseph F. Hennawi}
\affil{Department of Physics, University of California, Broida Hall, Santa Barbara, Santa Barbara, CA 93106-9530, USA}
\affil{Leiden Observatory, Leiden University, P.O. Box 9513, 2300 RA, Leiden, The Netherlands}
\author[0000-0002-3952-8588]{Leindert Boogaard}
\affil{Max-Planck-Institut f\"ur Astronomie, K\"onigstuhl 17, D-69117 Heidelberg, Germany}
\author{Anna Feltre}
\affil{INAF -- Osservatorio Astrofisico di Arcetri, Largo Enrico Fermi 5, I-50125 Firenze, Italy}
\author[0000-0003-2804-0648]{Themiya Nanayakkara}
\affil{Centre for Astrophysics and Supercomputing, Swinburne University of Technology, P.O. Box 218, Hawthorn, 3122, VIC, Australia}
\author{Roland Bacon}
\affil{Univ Lyon, Univ Lyon1, ENS de Lyon, CNRS, Centre de Recherche Astrophysique de Lyon UMR5574, 69230, Saint-Genis-Laval, France}
\author{Amy Barger}
\affiliation{Department of Astronomy, University of Wisconsin-Madison, 475 N. Charter St., Madison, WI 53706, USA}
\affiliation{Department of Physics and Astronomy, University of Hawaii,
2505 Correa Road, Honolulu, HI 96822, USA}
\affiliation{Institute for Astronomy, University of Hawaii, 2680 Woodlawn Drive,
Honolulu, HI 96822, USA}
\author[0000-0003-4359-8797]{Jarle Brinchmann}
\affil{Instituto de Astrof{\'\i}sica e Ci{\^e}ncias do Espa\c{c}o, Universidade do Porto, CAUP, Rua das Estrelas, PT4150-762 Porto, Portugal}
\author[0000-0002-8871-3026]{Marijn Franx}
\affil{Leiden Observatory, Leiden University, P.O. Box 9513, 2300 RA, Leiden, The Netherlands}
\author[0000-0002-0898-4038]{Takuya Hashimoto}
\affiliation{Division of Physics, Faculty of Pure and Applied Sciences, University of Tsukuba,Tsukuba, Ibaraki 305-8571, Japan}
\affiliation{Tomonaga Center for the History of the Universe (TCHoU), Faculty of Pure and Applied Sciences, University of Tsukuba, Tsukuba, Ibaraki 305-8571, Japan}
\author{Hanae Inami}
\affil{Hiroshima Astrophysical Science Center, Hiroshima University, 1-3-1 Kagamiyama, Higashi-Hiroshima, Hiroshima 739-8526, Japan}
\author{Haruka Kusakabe}
\affil{Observatoire de Gen\`eve, Universit\'e de Gen\`eve, 51 Ch. des Maillettes, 1290 Versoix, Switzerland}
\author{Floriane Leclercq}
\affil{Department of Astronomy, University of Texas at Austin, 2515 Speedway Blvd Stop C1400, Austin, TX 78712, USA}
\author{Lucie Rowland}
\affil{Leiden Observatory, Leiden University, P.O. Box 9513, 2300 RA, Leiden, The Netherlands}
\author[0000-0003-1282-7454]{Anthony J. Taylor}
\affil{Department of Astronomy, University of Texas at Austin, 2515 Speedway Blvd Stop C1400, Austin, TX 78712, USA}
\author{Christy Tremonti}
\affiliation{Department of Astronomy, University of Wisconsin-Madison, 475 N. Charter St., Madison, WI 53706, USA}
\author{Tanya Urrutia}
\affil{Leibniz-Institut f\"ur Astrophysik Potsdam (AIP), An der Sternwarte 16, D-14482 Potsdam, Germany}
\author[0000-0002-0668-5560]{Joop Schaye}
\affil{Leiden Observatory, Leiden University, P.O. Box 9513, 2300 RA, Leiden, The Netherlands}
\author{Charlotte Simmonds}
\affil{The Kavli Institute for Cosmology (KICC), University of Cambridge, Madingley Road, Cambridge, CB3 0HA, UK}
\affil{Cavendish Laboratory, University of Cambridge, 19 JJ Thomson Avenue, Cambridge, CB3 0HE, UK}
\author[0000-0001-5121-1260]{Elo\"ise Vitte}
\affil{European Southern Observatory, Alonso de Cordova 3107, Vitacura, Santiago, Chile}
\affil{Observatoire de Gen\`eve, Universit\'e de Gen\`eve, 51 Ch. des Maillettes, 1290 Versoix, Switzerland}

\begin{abstract}

Deep VLT/MUSE optical integral field spectroscopy has recently revealed an abundant population of ultra-faint galaxies ($M_{UV} \approx -15$; 0.01 $L_{\star}$) at $z=$2.9$-$6.7 due to their strong Lyman-$\alpha$ emission with no detectable continuum.  The implied \lya equivalent widths can be in excess of 100-200 \AA, challenging existing models of normal star formation and indicating extremely young ages, small stellar masses, and a very low amount of metal enrichment. We use JWST/NIRSpec's microshutter array to follow-up 45 of these galaxies (11h in G235M/F170LP and 7h in G395M/F290LP), as well as 45 lower-equivalent width Lyman-$\alpha$ emitters.  Our spectroscopy covers the range 1.7$-$5.1 micron in order to target strong optical emission lines: \ha, \oiii, \hb, and \nii. Individual measurements as well as stacks reveal line ratios consistent with a metal poor nature (2$-$40\% $Z_{\odot}$, depending on the calibration).  The galaxies with the highest equivalent widths of \lyans, in excess of 90 \AA, have lower\nii/\ha\ (1.9-$\sigma$) and \oiii/\hb\ (2.2-$\sigma$) ratios than those with lower equivalent widths, implying lower gas-phase metallicities at a combined significance of 2.4-$\sigma$.  This implies a selection based on \lyans\ equivalent width is an efficient technique for identifying younger, less chemically enriched systems.  

\end{abstract}

\section{Introduction}

Observing primordial galaxies has long been a goal in studies of the distant Universe.  However, the identification and subsequent follow-up of these sources present unique challenges.  Deep imaging surveys have efficiently identified galaxies at high-$z$ based on their rest-UV continuum, yet practical considerations due to the finite depth of the imaging mean that the very youngest sources remain undetected until they have built up sufficient stellar mass (and hence also metals).  Before this point, though, they would appear extremely luminous in Lyman-$\alpha$ $\lambda$1216 emission as their young, metal-poor stellar populations would be efficient producers of ionizing photons \citep{1967ApJ...148..377P}.  Indeed, theoretical predictions for metal-poor or metal-free stellar populations suggest that the strength of Lyman-$\alpha$ with respect to the UV continuum (i.e.\ the equivalent width; EW) would be far in excess of what can be observed in metal-enriched systems, with EW $\approx$ 200$-$240 \AA\ demarcating these populations \citep{1991ApJ...378..471C,2003AA...397..527S}.  Identifying sources with Lyman-$\alpha$ equivalent widths in excess of 200 \AA\ is a direct way to isolate galaxies before they have experienced significant metal enrichment \citep{2002ApJ...565L..71M, 2010AA...523A..64R, 2011MNRAS.418L.104Z}.

The MUSE integral field spectrograph on the VLT \citep{2010SPIE.7735E..08B} has opened a new parameter space for these searches: even in areas covered by the deepest optical/near-IR imaging in the \textit{Hubble} Ultra Deep Field \cite[UDF, $m_{\mathrm{AB}}$ $\approx$ 30;][]{2013ApJS..209....6I,2015AJ....150...31R}, 10$-$30 hour MUSE spectroscopy has identified numerous emission line galaxies without a detected continuum counterpart \citep{Paper1, 2017AA...608A..10H}.  In particular, MUSE has uncovered an abundant population of $z = 2.9 - 6.7$ high-EW Lyman-$\alpha$ emitters (LAEs) that are the faintest unlensed spectroscopically-confirmed galaxies ever detected at high redshift.  Detailed stacking analyses reveal the average UV continua, which are as faint as M$_{\mathrm{UV}} =$ -15 \cite[or 0.01 L$_{\star}$ at $z=4.5$;][]{2018ApJ...865L...1M}.  These LAEs, which represent 20\% of all detected LAEs in the field and 10\% of all expected galaxies at this $M_{\mathrm{UV}}$ \citep{2014ApJ...793..115B, Paper2}, must have high equivalent widths with some estimates exceeding the canonical 200$-$240 \AA$~$(rest-frame).  

\citet{2020MNRAS.493.5120M} stack 200h-deep \textit{Spitzer}/IRAC photometry for 35 of these sources at $z=4-5$, finding evidence for high-EW H$\alpha$ emission, indicating rapid stellar mass buildup.  They show that these sources are among the most efficient producers of ionizing photons ever discovered \cite[see also][]{2018ApJ...859...84H,2023arXiv230307931S}, and they are likely to be younger than 3 Myr and have gas-phase metallicities of 3$-$30\% Z$_{\odot}$.  However, these quantities are uncertain and constrained indirectly from the broadband HST and \textit{Spitzer} imaging data.

With the launch of JWST, rest-frame optical emission lines such as \oiii, \hb, and \ha\ have become directly observable beyond $z\approx2-3$, the limit for ground-based near-IR spectrographs.  Here we target some of the highest-EW LAEs with JWST/NIRSpec in order to detect these rest-frame optical emission lines.  The observed strengths and ratios of emission lines can be used to determine numerous physical quantities in these galaxies such as their star formation rates, gas-phase metallic abundances, and ionization properties, as has already been shown with Cycle 1 data from JWST \cite[e.g.\ ][]{2022ApJ...939L...3T,2022arXiv220807467B,2022arXiv221108255M,2022arXiv221015639R,2022AA...665L...4S,2022arXiv221204476W,2023arXiv230204298C,2023arXiv230109482F,2023arXiv230112825N,2023arXiv230106696S}.  In this first paper we focus on the ratio of \oiii\ $\lambda$5007 to \hb\ $\lambda$4861, with supplementary information provided by the ratio of \nii\ $\lambda$6584 to \ha\ $\lambda$6563.  These ratios use closely-spaced lines, providing resilience against the effects of dust attenuation and/or poorly flux-calibrated data.  Moreover, they can provide insight into the metal abundance of the gas in these galaxies, which are expected to be extremely low in the highest \lyans-EW systems.

The outline of this paper is as follows.  In Section \ref{sec:data} we present our sample and our JWST/NIRSpec observations.  In Section \ref{sec:results} we present the results of our spectroscopy, including emission line ratios from individual objects as well as stacks based on the EW of \lyans.  In Section \ref{sec:discussion} we discuss the implications of these results, make concluding remarks, and present prospects for future work with NIRSpec.

\section{Data}
\label{sec:data}

\subsection{Target Selection}

Our targets are selected from the MUSE HUDF survey \citep{2023AA...670A...4B} and the MUSE Wide survey \citep{2017arXiv170508215H, 2019AA...624A.141U}.  We preferentially target LAEs with plausibly high \lya equivalent widths, via bright \lya emission and/or faint measurements/limits on the rest-UV continuum magnitudes from the HST imaging \cite[e.g.\ ][]{2017AA...608A..10H,2018ApJ...865L...1M,2020MNRAS.493.5120M,2022AA...659A.183K}.

We prioritize sources based on their \lya EW and the likelihood that their rest-optical emission lines (\ha, \oiii, and/or \hb) are measurable with NIRSpec in the combination of G235M and G395M (see Section \ref{sec:obs}).  Specifically, we perform a first mask design with the NIRSpec Mask Planning Tool \cite[MPT;][]{2014SPIE.9149E..1ZK} at our assigned roll angle with the highest-EW sources only.  We then calculate the predicted wavelengths covered for each object and down-weight cases where the detector gap and/or the red edge of the NRS2 detector restricts the observability of these emission lines, based on the MUSE \lyans-based redshifts.  

We allow for spectra to overlap on the detector to maximize the multiplexing, with the understanding that this can lead to potential confusion in the 2D frames.  As our sources are emission line dominated \citep{2020MNRAS.493.5120M}, spatially compact \citep{2018ApJ...865L...1M}, and have known redshifts from MUSE spectroscopy, we can always dis-entangle the spectra on the detector provided the sources themselves are not overlapping in the cross-dispersion direction.  We impose a limit of two additional spectra overlapping with the spectra of any of our primary targets at any one position on the detector to minimize unnecessarily high background levels (a detailed description for how spectral overlap can be calculated as a function of shutter position will be given by N. Bonaventura et al. in prep.).

Our final microshutter array (MSA) design consists of $1\times3$ microshutters targeting 125 unique galaxies.  Of the 125, 64 are selected from the MUSE UDF DR2 catalog, 26 from the MUSE-Wide Survey, and 35 based on grism data from the 3D-HST survey \citep{2016ApJS..225...27M}.  For the following analysis, we only consider the galaxies that were selected based on the presence of \lya in their spectrum, i.e. the 90 galaxies selected from MUSE.  These galaxies have redshifts between 2.9 and 6.5, with a median redshift of 4.5.

\subsection{NIRSpec Observations}
\label{sec:obs}
Our MSA configuration was observed on 2022-09-11 (GO 1671).  Target acquisition was performed with MSATA using eight reference objects in three quadrants, the CLEAR filter, and NRSRAPID readout.  The HUDF does not contain a sufficient number of stars to use for the target acquisition process, so we supplement the reference target list with compact galaxies.  We utilize the ``stellarity'' measurements from the 3D-HST photometric catalog \citep{2014ApJS..214...24S} to pre-select compact objects.  We then estimate their magnitudes in the NIRSpec imaging filters (CLEAR, F110W, F140X) by fitting the 3D-HST photometry, including \textit{Spitzer}/IRAC 3.6 and 4.5 micron data, using \texttt{MAGPHYS} \citep{2008MNRAS.388.1595D} at the best-fit grism redshift \citep{2016ApJS..225...27M}.  The MPT selects the eight objects from this larger list for the target acquisition.

The MSA configuration was observed using a three-point nodding pattern.  Each nodded exposure consisted of three integrations: 20 groups per integration with NRSIRS2 in G235M/F170LP and 100 groups per integration with NRSIRS2RAPID in G395M/F290LP\footnote{The different readout patterns result in different data rates, the limitations of which prevent us from observing both in the preferred NRSIRS2RAPID.}.  With three total nodding sequences in G235M/F170LP, we obtain a total of 39,783.9s of exposure time; with two total nodding sequences in G395M/F290LP, we obtain a total of 26,522.6s of exposure time.  Only one of our primary MUSE-selected targets was affected by an inoperable microshutter (see Appendix \ref{sec:appendix_msa}).

The NIRSpec data was reduced using level 1 and level 2 of the version 1.10.2 NIRSpec pipeline and the \texttt{jwst\_1097.pmap} context \citep{bushouse_howard_2022_7229890}.  We perform an additional $1/f$ noise correction and a median bias correction using \texttt{msaexp}\footnote{\url{https://github.com/gbrammer/msaexp/releases/tag/0.3.4}}.
 Calibrated, un-rectified 2D frames are combined using \texttt{PypeIt}\footnote{\url{https://pypeit.readthedocs.io/en/latest/}} \citep{pypeit:joss_pub, pypeit:zenodo}.  The sky background, bar shadows, and detector-level artifacts are removed with a pixel-level background subtraction based on adjacent exposures in the nodding sequence.

 We note that in this version of the NIRSpec pipeline, the errors are systematically under-estimated compared to the measured pixel-to-pixel variation in the fluxes \cite[as seen in e.g.\ ][]{2022arXiv220712388T,2022arXiv220807467B}.  We observe this already in the level 1 rate files, where the observed pixel-to-pixel flux variations in two consecutive exposures are larger than the quoted uncertainties: in the case of G235M (NRSIRS2 readout) this is a factor of 1.91, and in the case of G395M (NRSIRS2RAPID readout) this is a factor of 1.70.  These discrepancies can be caused by a number of issues that are currently not being accounted for in the pipeline, such as thermal instabilities or inter-pixel capacitance.  For what follows, we include these empirical factors in all of our uncertainties; a more detailed discussion can be found in Appendix \ref{sec:appendix_noise}.  While this has a small impact on what we consider to be a detection, there is no evidence that this would have a systematic impact on the stacked line ratios that are the main result of this paper.

\begin{figure*}
    \centering
    \includegraphics[width=0.95\textwidth]{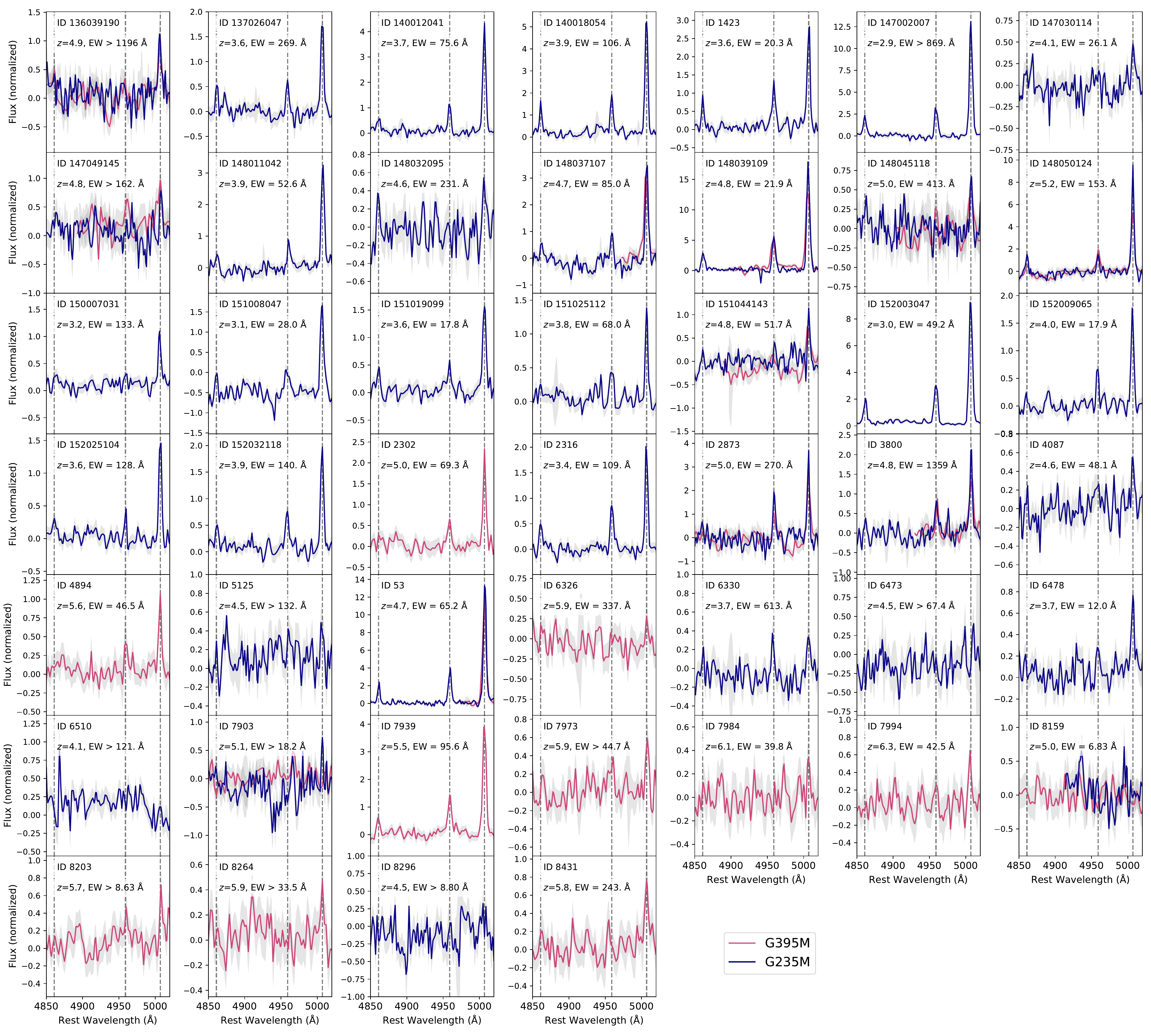} %pub_plot_all.py
    %\vspace{-1cm}
    \caption{NIRSpec spectra of LAEs showing the region around \oiii\ and \hb for objects in our sample.  The darker (blue) line shows G235M data and the lighter (pink) line shows G395M data.  Shaded regions show the $\pm$1-$\sigma$ errors on the spectra.}
    \label{fig:specs}
\end{figure*}

\begin{figure*}
    \centering
    \includegraphics[width=0.95\textwidth]{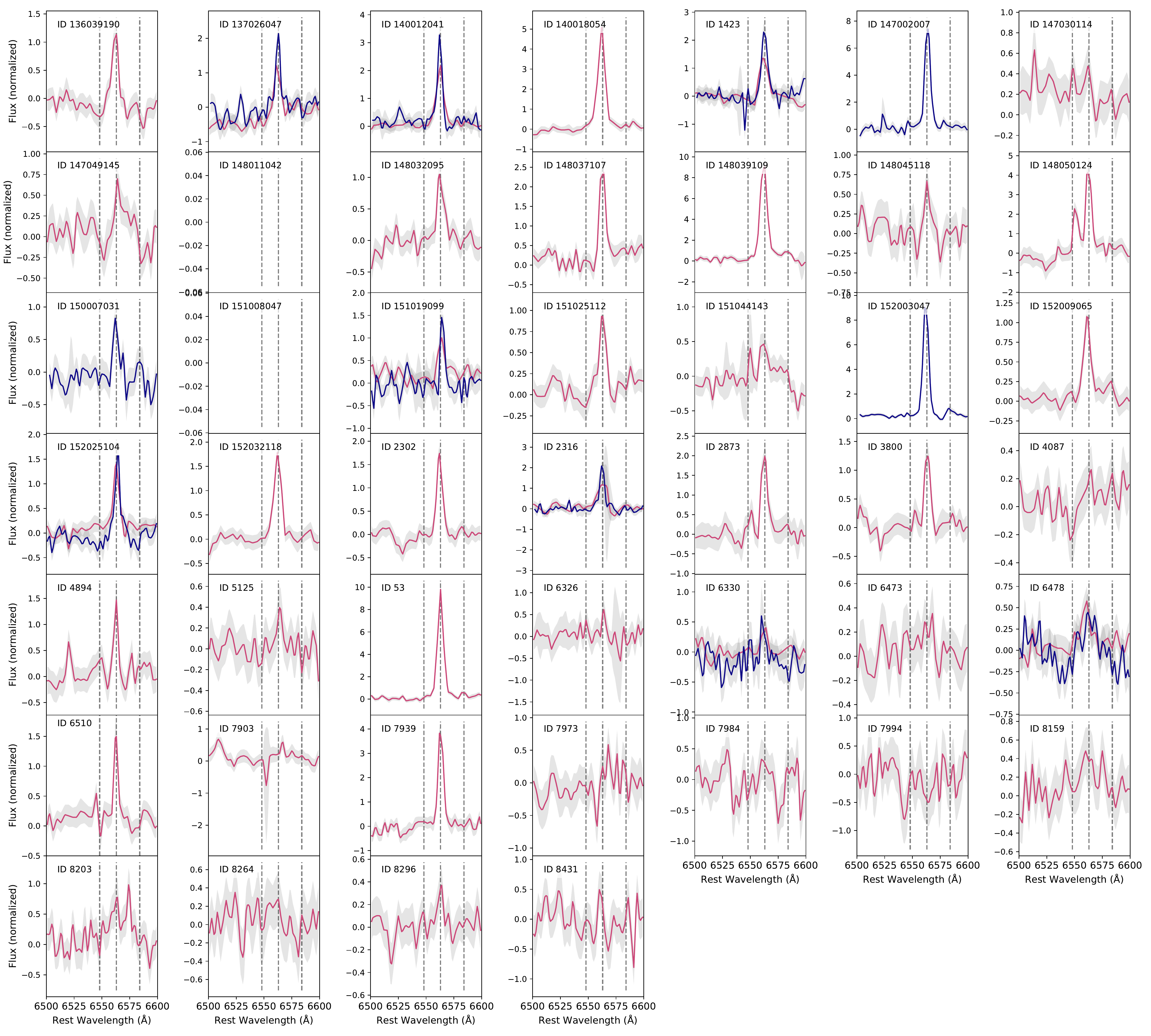} %pub_plot_all.py
    %\vspace{-1cm}
    \caption{Same as Figure \ref{fig:specs}, but for the region around \nii\ and \ha.}
    \label{fig:specs2}
\end{figure*}

\subsection{Spectral Extractions and Redshift determinations}
\label{sec:spec}

Our observed sample consists of 90 LAEs selected from MUSE spectroscopy.  Although the peak of \lya emission is typically known to be offset from the systemic redshift of the galaxy by several hundred km s$^{-1}$ \cite[e.g.\ ][]{2003ApJ...588...65S,2015ApJ...809...89T,2020MNRAS.496.1013M,2021MNRAS.505.1382M}, redshifts from the peak of the \lya emission in MUSE nonetheless give us an idea of where to look for rest-optical emission in the NIRSpec data.  In future work we will systematically assess the incidence rate of emission as a function of source properties.  

We begin by extracting a cutout of the reduced, unrectified 2D spectrum at the expected position of $\lambda_{\mathrm{H\alpha}} \pm$ 50 \AA, and between $\lambda_{\mathrm{H\beta}} - 50$ \AA\ and $\lambda_{\mathrm{[OIII] 5007}} + 50$ \AA\ ($\Delta\lambda \approx$ 1000 km s$^{-1}$), where $\lambda$ represents the predicted line centers based on the \lya redshifts.  Within these 2D cutouts, we visually inspect for the presence of strong emission lines (including negative residuals from the nodded background subtraction) and remove spurious cases where e.g.\  a strong cosmic ray residual with under-estimated uncertainties can mimic a true emission line.  The widths of the search windows are narrow enough that we would not confuse e.g. \hb\ and \oiii\ in the case that only a single line is present in the spectrum.

1D spectra are extracted using a boxcar aperture with a size corresponding to the nodding length (0\farcs5) that is constant with wavelength.  In cases where a line is visible in the 2D spectrum, we extract the boxcar at the spatial position of the strongest line, taking into account the curvature of the slits.  Otherwise, we center the boxcar on the catalog source position for each object (i.e. its RA/Dec as projected onto the MSA shutter), meaning the position that corresponds to the flux-weighted \lya centroid from MUSE.

When a line is observed in the NIRSpec data, we use the wavelength corresponding to the peak line flux as the initial guess to the redshift fitting on the 1D spectrum with \pf.  Otherwise, we start the \pf\ fit at the MUSE-based \lya redshift, but allowing for alternative redshift solutions to be found (see \citeauthor{2023AA...670A...4B} \citeyear{2023AA...670A...4B} for a description of the \pf\ redshift fitting process, and Appendix \ref{sec:appendix_z} for a description of the accuracy of the NIRSpec wavelength calibration).  Our final sample of MUSE LAEs with redshifts from our NIRSpec observations consists of 46 objects (i.e. 46/90 objects from our initial sample have at least one emission line detected).

Figures \ref{fig:specs} and \ref{fig:specs2} show the NIRSpec spectra for all objects in the sample with at least one detected optical emission line.  The objects for which we did not detect any rest-optical emission lines with NIRSpec are either too faint, are affected by artefacts in the data, and/or have an incorrect MUSE-based redshift.  Stacked NIRSpec spectra for undetected objects based on the MUSE redshifts may shed light on the relative contributions of each failure mode and will be presented in future work.

\subsection{Line flux determinations}
\label{sec:flux}

Once the redshift for each spectrum is fixed, we perform a Markov Chain Monte Carlo analysis of the spectra using \pf.  Spectra are perturbed according to their uncertainties and re-fit until convergence, with the resulting mean and standard deviation of the line flux measurements being quoted throughout.  We fit the G235M and G395M data separately. In cases when a line is covered in both we quote the G235M result as those observations have longer exposure times and hence a higher expected S/N.  To minimize compounding assumptions about the absolute error values as well as the flux calibrations (see Appendix \ref{sec:systematics}), we do not attempt to combine the spectra in these overlapping regions.

As an additional verification of our uncertainties, we compare the measurements for emission lines from each of the individual three-nodded sequences described in Section \ref{sec:obs}.  As discussed in \citet{2022arXiv220807467B}, these ``duplicate'' observations are useful in understanding systematic issues with the uncertainties such as the impact of correlated noise (which e.g.\  would not be apparent when looking at pixel-to-pixel variations alone as in Appendix \ref{sec:appendix_noise}).  We use our line fitting methodology on each of the individual exposures and compare the measurements and the uncertainties for \oiii, \hb, and \ha.  The standard deviation of the resulting flux differences is 0.996, justifying our increase of the pixel-level errors in Section \ref{sec:obs}.

\section{Results}
\label{sec:results}

For brevity, we define the following emission line ratios:\\

$\mathrm{N2 = \text{log ([N {\sc ii}]} \lambda 6584 / \text{H} \alpha)}$\\
    
$\mathrm{R3 = \text{log ([O {\sc iii}]} \lambda 5007 / \text{H} \beta)}$.\\

Note that the ratios of \oiii\ $\lambda$5007 to $\lambda$4959 and \nii\ $\lambda6584$ to $\lambda6548$ are fixed in our fitting procedure to the values in \citet{2000MNRAS.312..813S}.

Considering the close wavelength spacing between the lines in both ratios, they are relatively insensitive to dust attenuation and/or wavelength-dependent flux calibration issues (see Appendix \ref{sec:systematics}).  For each galaxy, we calculate the ratio according to the measured fluxes and uncertainties of each emission line as detailed in Section \ref{sec:spec}.  In the case where one of the emission lines is not detected above a signal-to-noise of 1, we instead use the 2-$\sigma$ upper limit to the flux when calculating the limit on the line ratio.

For the 44 galaxies with spectral coverage of both N2 and R3 ratios, we can construct line ratio diagnostic diagrams which can be used to differentiate between star-formation and active galactic nucleus ionization in the galaxy \cite[e.g.\ ][]{1981PASP...93....5B,1987ApJS...63..295V, 2001ApJ...556..121K}.  All of our objects with measurements of both ratios are consistent with being in the star-forming region using the demarcation line of \citet{2013ApJ...774..100K} extrapolated to the median redshift of our sample $z=4.5$ within 1-$\sigma$.  Beyond this simple demarcation, though, these line ratios encode information about the interstellar medium (ISM) conditions inside the galaxies.  In particular, galaxies with different ionizing radiation fields, metallicities, electron densities, and ionization parameters will lie in different parts of line ratio space \citep{2013ApJ...774..100K}.

\begin{figure*}
    \centering
    \includegraphics[width=0.95\textwidth]{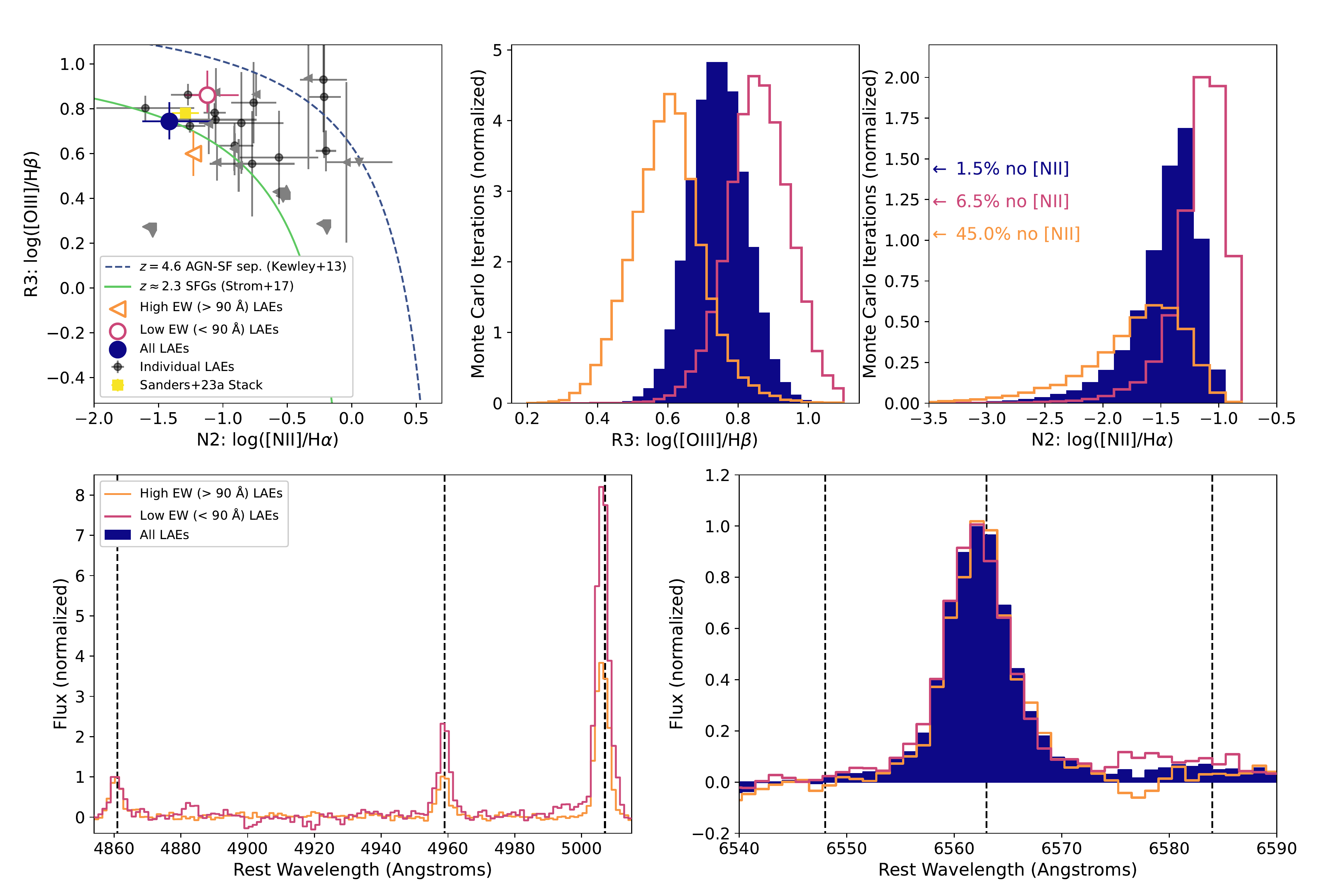} %stack_NIRSpec_fortalk2.py
    \caption{(Upper left) R3 versus N2 ionization diagnostic plot for our sample of LAEs at $z\approx4.6$.  For objects with S/N $<$ 1 in any of the individual lines, we use gray triangles to represent limits (2-$\sigma$) on the ratio(s); objects without a S/N $>$ 1 detection in both components of R3 or N2 are omitted, although they are still included in the stacks.  The stacked values from our full sample of LAEs lie on the sequence of $z\approx2.3$ star-forming galaxies (SFGs) from \citet{2017ApJ...836..164S} and below the $z=4.6$ extrapolated demarcation between active galactic nuclei (AGN) and star formation from \citeauthor{2013ApJ...774..100K} (\citeyear{2013ApJ...774..100K}, although we caution that this extrapolation is not well-calibrated at these redshifts).  Compared
to the \citet{2023arXiv230106696S} stack of $z\approx4.5$ SFGs, we observe a trend of lower R3 and N2 values when restricting the stack to higher-EW \lyans.
    (Upper center and right) Observed R3 and N2 ratios for our sample from each Monte Carlo iteration (Section \ref{sec:stacks}).  (Lower panels) Stacked spectra for the three EW subsamples showing \hb, \oiii, \ha, and \nii\ (vertical dashed lines).  Each plot is normalized to the flux of \hb\ or \ha.}
    \label{fig:lineratios}
\end{figure*}

In Figure \ref{fig:lineratios} we present our measurements as well as limits on the line ratios of R3 and N2 for our sample of MUSE-selected LAEs.  36 of our 46 targets have detections or upper/lower limits on R3 and N2.  Similar to the $z>4$ star-forming galaxies from \citet{2023arXiv230204298C}, we see relatively few detections (19) of \nii\ in our individual spectra.  Particularly at the typical \oiii/\hb\ ratios observed here, the lack of \nii\ limits our ability to interpret the physical conditions of the ISM in individual galaxies.  A more detailed investigation of these conditions utilizing additional line ratios will be presented in forthcoming work (Z. Lewis et al. in prep), but here we focus on stacked spectra where precise measurements of the (average) line ratios are possible.

\subsection{Line ratios from stacked spectra}
\label{sec:stacks}

In order to achieve a higher S/N in each of the lines, we create median stacks of our NIRSpec data.  For the 46 MUSE LAEs with confirmed redshifts from NIRSpec, we split our sample into two parts based on their \lya EWs.  Note that the stacks can include objects that are missing spectral coverage of either R3 or N2 and otherwise do not appear in Figure \ref{fig:lineratios}.  We select 23 objects that (a) have a measured \lya rest-frame EW in excess of 90 \AA\ or (b) are undetected in the HST imaging, implying we can only derive a lower-limit on the \lya EW \citep{2020MNRAS.493.5120M,2022AA...659A.183K}.  This constitutes our ``high EW'' stack, with a median EW of 154 \AA.  We create a second stack including the 23 objects with lower equivalent widths (median value of 43 \AA, cf. the mean \lya EW from MUSE-detected sources at $z=4.9$ of 88 \AA; \citeauthor{2017AA...608A..10H} \citeyear{2017AA...608A..10H}). The final stack consists of all 46 MUSE LAEs with redshift detections in NIRSpec, noting that all of these sources are well-detected in \lya from MUSE.  The median redshift for the stacks are $z=4.8, 4.8,$ and $4.6$, respectively. Spectra in each bin are interpolated onto the same rest-frame wavelength grid and the median flux at each wavelength is measured.  While the individual measurements of emission line ratios can be quite noisy, typically from very faint \nii\ or \hb, the stacked measurements result in significant ($>$2-$\sigma$) detections in all emission lines.  Error bars are determined by taking 50,000 random samples of objects in each of the categories and re-performing the median stacking including random perturbations to each 1D spectrum according to its noise vector.  

Figure \ref{fig:lineratios} shows the resulting stacks (lower panels) as well as the values for R3 and N2 from each of the individual Monte Carlo iterations (upper-center and upper-right panels) and the overall R3 and N2 values for each stack (upper-left panel).  In the case of N2 for the high-EW stack, we only obtain a reliable upper limit to the value as \nii\ is not always detected in the stacks (45.0\% of the time the \nii\ flux is either zero or negative).  The stack of all LAEs has comparable line ratios to the \citet{2023arXiv230106696S} stack of star forming galaxies at similar redshifts (not selected by \lyans), which itself closely resembles the values for $z\approx2.3$ star forming galaxies from \citet{2017ApJ...836..164S}.

\begin{figure*}
    \centering
    \includegraphics[width=0.95\textwidth]{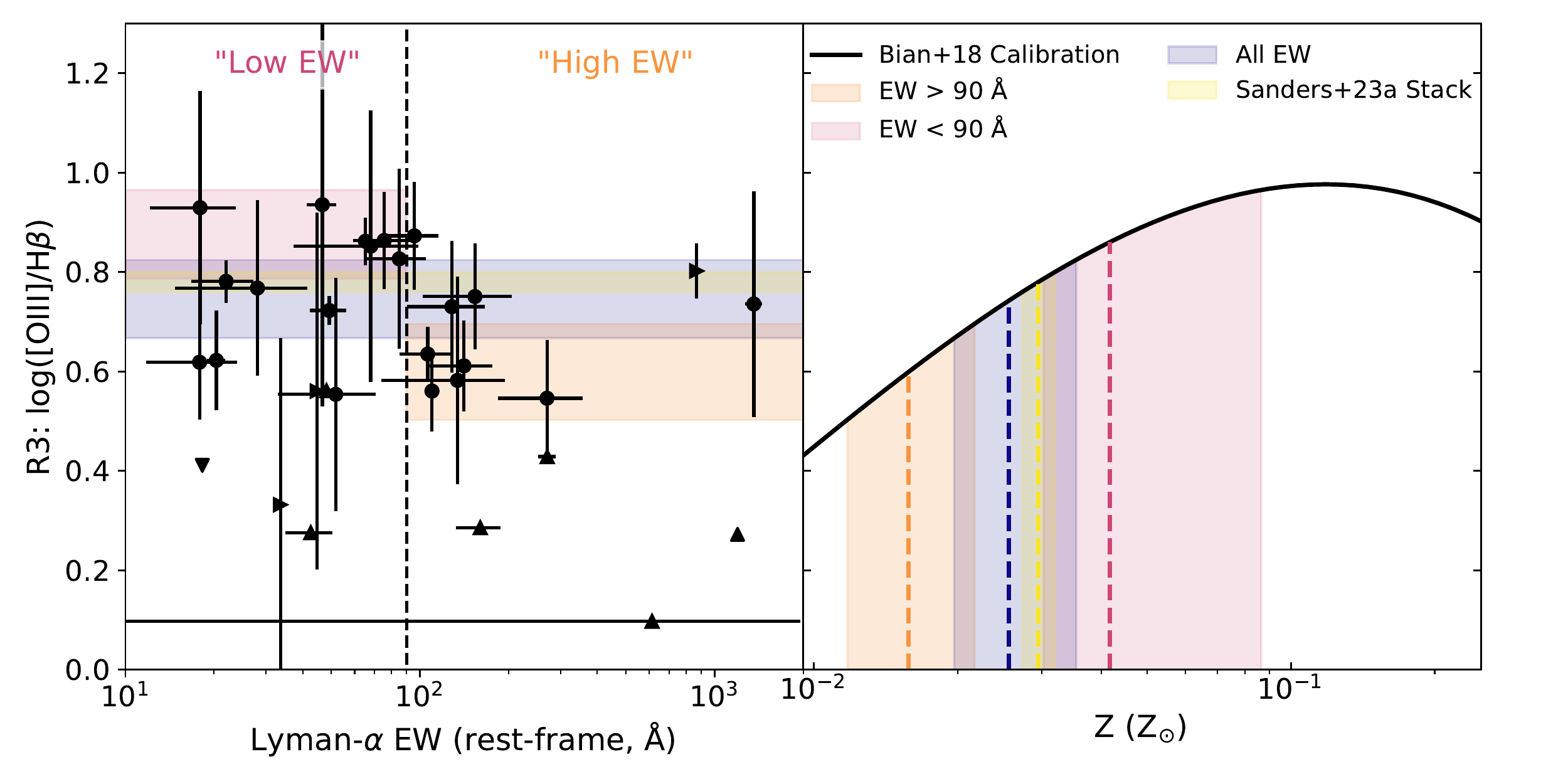} %stack_NIRSpec_fortalk2.py
    \caption{(left) NIRSpec-derived R3 ratio versus MUSE-derived \lya EW.  In cases where the \lya EW or R3 value are limits, triangles are used.  The shaded regions are the $\pm$1-$\sigma$ ranges for R3 for each of the stacks based on \lya EW. We observe systematically lower R3 values as a function of \lya EW, indicative of lower gas-phase metallicities. (right) The relationship between R3 and gas-phase metallicity using the calibration from \citet{2018ApJ...859..175B}.  The shaded regions show the 1-$\sigma$ stacked R3 ranges, including those from \citet{2023arXiv230106696S}, and their corresponding metallicities. Our high-EW LAEs have lower R3 values than the ``typical'' SFGs from \citet{2023arXiv230106696S} at a significance of 2.0-$\sigma$, indicating lower metallicities (vertical dashed lines).}
    \label{fig:lya-r3}
\end{figure*}

Within our sample, we see lower R3 and N2 values in the stack of sources with the highest \lya EW, implying a lower gas-phase metallicity.  Using the Monte Carlo iterations, we see that N2 is lower in the high-EW stack at a significance of 1.9-$\sigma$ compared to the low-EW stack.  R3 is lower at a significance of 2.2-$\sigma$.  Figure \ref{fig:lya-r3} shows the relationship between R3 and \lya EW, providing further evidence that higher EWs correspond to lower R3 values.  This strongly suggests a lower gas-phase oxygen abundance in the high-EW LAEs, as we discuss in the following Section.  Furthermore, since the lines are closely spaced in wavelength it is a robust result that is not strongly affected by systematics (see Appendix \ref{sec:systematics}).

\section{Discussion and Conclusions}
\label{sec:discussion}

As shown in \citet{2016ApJ...832..171T,2018ApJ...865L...1M, 2020MNRAS.493.5120M} and \citet{2022AA...659A.183K}, an elevated \lya EW typically signifies a low-metallicity, young, vigorously star-forming galaxy.  This trend of higher \lya EWs with lower gas phase metallicities typically manifests only as a lower N2 value; the R3 values are typically \textit{elevated} in the higher-EW systems \citep{2016ApJ...832..171T,2020MNRAS.495.1501C,2021ApJ...920...95D}.  This, however, is expected for metallicities in excess of $\approx$10\% \zsol\ (seen in the right panel of Figure \ref{fig:lya-r3}) where R3 increases as the metallicity decreases.  At the extremely low metallicities of the MUSE-selected sources, we expect metallicity to be directly correlated to the value of R3.  This difference is due to the double-valued nature of R3.

Although Nitrogen-based calibrations such as R3N2 (i.e. the ratio of R3 to N2) have not been calibrated at these redshifts using JWST data \citep{2023arXiv230603120L}, if we were to adopt the R3N2 calibration of \citet{2018ApJ...859..175B} we would observe a lower metallicity in the high-EW subset compared to the low-EW subset at a significance of 2.39-$\sigma$.  If we use the empirical calibration between R3 and oxygen abundance from \citet{2018ApJ...859..175B}, which has been shown to match high-$z$ observations with NIRSpec \citep{2023arXiv230603120L}, we observe that the highest-EW objects have a metallicity of 1.5$\pm$0.50 \% \zsol, compared to 4.1$_{-1.1}^{+4.5}$ \% \zsol\ for the lower-EW subset and 2.5$_{-0.60}^{+0.98}$ \% \zsol\ for all LAEs (Figure \ref{fig:lya-r3}).  If we use the nebular photoionization models from \citeauthor{2016MNRAS.462.1757G} (\citeyear{2016MNRAS.462.1757G}, which assume a constant star formation rate for the last 100 Myr) to match the R3 and N2 values of our low- and high-EW samples, we determine oxygen abundances of 60$_{-22}^{+10}\%$ and $< 37\%$ \zsol\ (2-$\sigma$), respectively.  The stack of all LAEs would have a metallicity of 20$_{-19}^{+11}$\% \zsol.

The discrepancy between the absolute metallicities derived using the two different methods partially highlights the need for strong-line diagnostics calibrated from JWST data that include Nitrogen lines \citep{2023arXiv230308149S,2023arXiv230603120L}.  We also need a detailed characterization of the star formation histories and emission line properties of individual galaxies in the sample (Z. Lewis in prep.).  However, all of these values are in the range of metallicities derived by \citet{2020MNRAS.493.5120M}, based on the \lya EWs, the ionizing photon production efficiencies, and the UV continuum slopes.

If higher \lya EWs indeed correspond to lower metallicities it would be expected that this MUSE-derived sample, with half of the sources plausibly having EWs in excess of 90 \AA, would show a lower metallicity than the \citet{2016ApJ...832..171T} sample with a median EW of 56 \AA.  This is what we observe, where the stack of all of our LAEs has a metallicity of 2.5\% \zsol\ (or 20\% \zsol\ using the \citeauthor{2016MNRAS.462.1757G} \citeyear{2016MNRAS.462.1757G} calibrations) compared to 22\% \zsol\ for the \citet{2016ApJ...832..171T} sample.  Moreover, \citet{2016ApJ...832..171T} observe that the continuum-faintest sources in their sample (i.e. those with plausibly the highest EWs) may actually be in the low-metallicity regime with a depressed value of R3.  This has a dependence on the ionization parameter, where \citet{2016ApJ...832..171T} observe a positive correlation between \lya EW and the ionization parameter.  In fact, a larger ionization parameter may actually drive the elevated \lya EWs observed in our sample \citep{2016ApJ...830...52E}.  While we cannot directly constrain the ionization parameter with the optical lines presented in this work, future stacking experiments with these data can potentially yield constraints.

JWST/NIRSpec has given us the unique opportunity to study the rest-frame-optical spectral region of galaxies at $z\approx4-6$ for the first time.  With 7 and 11 hours of exposure time in G395M/F290LP and G235M/F170LP, respectively, we detect \oiii, \hb, \ha, and/or \nii\ in 46 out of 90 MUSE-selected Lyman-$\alpha$ emitters down to a UV magnitude of $M_{\mathrm{UV}}\approx-15$.  We measure the \nii/\ha\ ratio (limits) for 19 (25) galaxies and the \oiii/\hb\ ratio for 25 (21) galaxies.  13 galaxies have detections of all four emission lines studied here.  After correcting the error estimates propagated through the data reduction pipeline (Appendix \ref{sec:appendix_noise}), we focus on these same line ratios in stacked spectra.  We observe that, when split into ``high'' and ``low'' Lyman-$\alpha$ EW stacks at 90 \AA, the ``high'' stack consists of galaxies with systematically lower gas-phase metallicities, potentially as low as 1.5\% \zsol when using the \citet{2018ApJ...859..175B} R3 calibration.  The ``low'' EW stack is still plausibly very low metallicity (4.1\% \zsol), with R3 and N2 values that are lower than similar results from other intermediate- to high-$z$ stacks.  We stress that stacking is important to take into account a number of galaxies that lack individual detections of \nii\ and/or \hb, where otherwise requiring detections of those lines would bias the sample mean to higher \oiii/\hb\ and \nii/\ha\ ratios.

In galaxies with sub-solar metallicity, one of the primary sources of radiative cooling is via collisionally excited ions like \oiii.  However, at oxygen abundances less than $\approx$10\% \zsol, there is so little oxygen in the ISM that the ratio of \oiii\ to \hb\ would start to decrease (see Figure \ref{fig:lya-r3}).  Although this has been observed in some of the most metal-poor galaxies locally \cite[e.g.\  ][]{2016ApJ...827..126B,2022ApJ...935..150L,2022ApJ...930...37U}, this is the first evidence so far of a systematically suppressed \oiii/\hb\ ratio at high $z$ from JWST spectroscopy \citep[e.g.\ ][]{2023arXiv230106696S,2023arXiv230308149S,2023arXiv230204298C,2023arXiv230112825N,2023arXiv230603120L}.  We interpret this as evidence for extremely low metallicities in the highest-EW LAEs.  This, however, is still a significant level of metal enrichment compared to truly ``Population III'' stellar populations which are expected to show R3 ratios below $-$1 \citep{2011MNRAS.415.2920I}.  
In future work we will utilize improved data reduction and analysis tools to derive metallicities and excitation diagnostics for individual galaxies.  With absolute flux calibrations, we will determine star formation rates from \hb\ to compare to the results of \citet{2020MNRAS.493.5120M}, who determined a nominal \ha-based SFR of 1.2 \msol\ yr$^{-1}$ and a UV-based SFR of 0.1 \msol\ yr$^{-1}$ in the highest EW LAEs (the core of this sample), where the discrepancy could be due to episodic star formation histories. Nevertheless, our initial results with JWST/NIRSpec already highlight the huge advances that can be made with this facility.

\begin{acknowledgments}
The authors would like to thank Peter Jakobsen, Allison Strom, Gwen Rudie, Ryan Trainor, and Gabe Brammer for numerous productive discussions about NIRSpec, as well as Rychard Bouwens, Thierry Contini, Josie Kerutt, Ivo Labb\'e, Roser Pello, Johan Richard, Kasper Schmidt, Anne Verhamme, and Lutz Wisotzki for work on the proposal.\\ 
This work is based on observations made with the NASA/ESA/CSA \emph{James Webb Space Telescope}. The data were obtained from the Mikulski Archive for Space Telescopes at the Space Telescope Science Institute, which is operated by the Association of Universities for Research in Astronomy, Inc., under NASA contract NAS 5-03127 for \JWST. MM acknowledges support from the National Science Foundation via AAG grant 2205519 and the Wisconsin Alumni Research Foundation via grant MSN251397.  ZL and CT acknowledge financial support from NASA via JWST-GO-1671. RB acknowledges support from the ANR L-INTENSE (ANR-20-CE92-0015). JB acknowledges financial support from the Funda\c{c}\~{a}o para a Ci\^{e}ncia e a Tecnologia (FCT) through national funds PTDC/FIS-AST/4862/2020, UIDB/04434/2020, UIDP/04434/2020, and work contract 2020.03379.CEECIND. TH is supported by Leading Initiative for Excellent Young Researchers, MEXT, Japan (HJH02007) and by JSPS KAKENHI Grant Numbers (20K22358 and 22H01258). HI acknowledges support from JSPS KAKENHI Grant Number JP19K23462 and JP21H01129.

\end{acknowledgments}

\facilities{JWST, VLT:Yepun}

\software{astropy \citep{2022ApJ...935..167A}; grizli \citep{brammer_gabriel_2023_7712834}; JWST Calibration Pipeline \citep{bushouse_howard_2022_7229890}; magphys \citep{2008MNRAS.388.1595D}; matplotlib \citep{Hunter:2007}; numpy \citep{harris2020array}; pandeia \citep{10.1117/12.2231768}; pypeit \citep{pypeit:joss_arXiv}; pyplatefit \citep{2023AA...670A...4B}}

\appendix

\section{NIRSPEC Microshutter Operability}
\label{sec:appendix_msa}
As noted in \citet{2022SPIE12180E..3RR} as well as papers on ERO data \cite[e.g.\ ][]{2022ApJ...939L...3T,2023MNRAS.518..425C}, there can be intermittent failures of otherwise functional MSA shutters.  We diagnose the incidence rate of this affect by comparing our planned MSA configuration with our post-acquisition confirmation image, which ultimately reveals which shutters are allowing flux transmission during the science exposures (nb. we cannot fully diagnose if any of these shutters are \textit{partially} open, but this would not affect our primary science as we are only concerned with line ratios and not any absolute flux calibration).

Of the 386 shutters that were commanded open in our observations, three did not open (0.8\%). Two of the failures were shutters in quadrant 2: (51, 7) and (151, 159); one was in quadrant 4: (91, 151). As of APT version 2022.7.1, none of these shutters are determined to be inoperable.  This is significantly smaller than the $\approx$4\% quoted by \citet{2022SPIE12180E..3RR}.  However, this could be due to the relatively few commanded-open shutters placed in quadrants 1 and 2 (90 and 48, respectively) compared to quadrants 3 and 4 (87 and 161), as shutters in these two quadrants are approximately twice as likely to experience an intermittent failure \citep{2022SPIE12180E..3RR}.

\section{NIRSpec Noise Estimates}
\label{sec:appendix_noise}
\begin{figure*}
       \centering
    \includegraphics[width=0.9\textwidth]{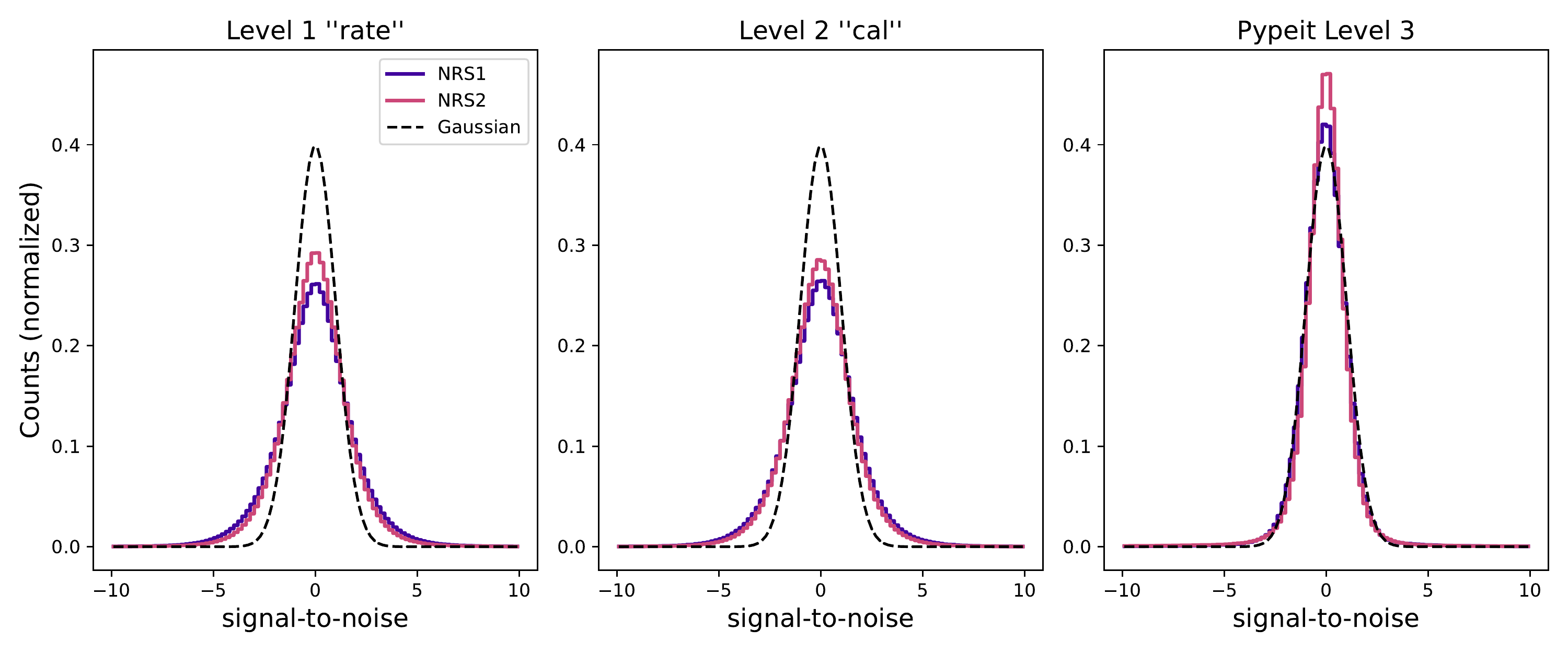}
    \caption{Comparison of the background-subtracted signal-to-noise values for (left) all pixels in the level 1 processed ``rate'' files, (center) pixels illuminated by open MSA shutters in the level 2 ``cal'' files, and (right) pixels in the \texttt{Pypeit}-calibrated 2D frames, including the extra factors of 1.91 and 1.70 in the noise estimates as described in the text.  Each of the two detectors are plotted separately.  The discrepancy in the distributions of the levels 1 and 2 products compared to Gaussian statistics (dashed line) motivate the extra noise factors used in \texttt{Pypeit}.}
    \label{fig:noise} 
\end{figure*}

Using the STScI NIRSpec pipeline as described in Section \ref{sec:obs}, the quoted uncertainties in the 2D rate files do not accurately explain the pixel-to-pixel variations present in the data: the errors are systematically under-estimated. In Figure \ref{fig:noise}, we show data from both the standard level 1 and 2 products (``rate'' and ``cal'' files) in the left and center panels, respectively. We plot the background-subtracted flux (differenced using our nodding pattern) compared to the quoted uncertainties on each of the pixels. If the noise is properly accounted for, this would be a Gaussian distribution with a $\sigma$ of 1. However, this is not observed to be the case (nb. without a median bias level subtraction performed by \texttt{msaexp}, the NRS2 values are offset towards negative values instead of being centered on zero, visible in Figure \ref{fig:noise}).  To account for this, we empirically determine a multiplicative factor that would turn these distributions in to normal Gaussians.  We therefore scale all of our error estimates by a factor of 1.91 (G235M) or 1.70 (G395M), leading to the distributions seen in the right panel of Figure \ref{fig:noise}.

\section{NIRSPEC WAVELENGTH CALIBRATION}
\label{sec:appendix_z}
Unlike ground-based spectrographs, there are typically no dedicated arc calibrations taken with NIRSpec after MSA observations, nor are there OH skylines that can be used.  As such, it is valuable to provide additional verification that the wavelength calibration step is performed correctly.  We have the opportunity to check this with our R1000 spectra in cases where the G235M and G395M gratings overlap (i.e. $\approx$28500$-$30000 \AA).

Results for individual objects are given in Table \ref{tab:tab}.  The mean velocity offset between the two gratings is 22.38$\pm$16.75 km s$^{-1}$, i.e. it is consistent with no systematic offset between the two gratings. 
\begin{deluxetable}{ccccc}
\tablecaption{Comparison of emission line redshifts in G235M and G395M}
\tablehead{
\colhead{MUSE ID} & \colhead{Line} & \colhead{$z_{G235M}$} & \colhead{$z_{G395M}$} & \colhead{$\Delta$v (km s$^{-1}$)}}
\startdata
53 & \oiii\ & 4.7763$\pm$5.9841e-5 & 4.7757$\pm$1.0410e-4 & 31.030$\pm$6.2364 \\
2316 & \ha\ & 3.4713$\pm$4.1071e-4 & 3.4702$\pm$1.6707e-3 & 72.638$\pm$115.43\\
2873 & \oiii\ & 5.0499$\pm$1.9912e-4 & 5.0500$\pm$3.2134e-4 & -8.4822$\pm$18.746 \\
3800 & \oiii\ & 4.8208$\pm$1.7813e-4 & 4.8214$\pm$6.0100e-4 & -32.666$\pm$36.234\\
136039190 & \oiii\ & 4.9438$\pm$9.9560e-4 & 4.9440$\pm$7.2192e-4 & -9.6310$\pm$62.070\\
140012041 & \ha\ & 3.7931$\pm$2.0336e-4 & 3.7932$\pm$2.7577e-4  & -2.7580$\pm$21.446\\
148039109 & \oiii\ & 4.8682$\pm$5.3343e-5 & 4.8674$\pm$7.9943e-5 & 38.600$\pm$4.9132\\
151044143 & \oiii\ & 4.8677$\pm$3.3596e-4 & 4.8665$\pm$9.4421e-4  & 61.655$\pm$51.240 \\
152025104 & \ha\ & 3.6585$\pm$1.9973e-4 & 3.6577$\pm$3.7174e-4  & 50.993$\pm$27.176
\enddata
\tablecomments{Redshift determinations are all from \pf, using a Monte Carlo algorithm to determine the uncertainties.  When multiple lines are available in both gratings (i.e. \oiii\ and \hb), we utilize the strongest line in the group for the redshift determination.\label{tab:tab}
}
\end{deluxetable}

\section{Systematic Uncertainties on \oiii/\hb\ and \nii/\ha}
\label{sec:systematics} 

Due to the fixed nature of the NIRSpec MSA shutters and the wavelength-dependent point spread function, any emission lines that are separated in wavelength will suffer from different amounts of geometrical slit losses.  For our input catalog, the NIRSpec pipeline currently assumes that all sources are all spatially extended and treats them as uniformly filling the open MSA area.  Since the geometrical slit losses will increase with wavelength, the pipeline correction could systematically bias our R3 and N2 values low. 
 
 Although these pairs of lines are close together in wavelength, we may nevertheless be systematically underestimating the line ratios due to this effect.  In order to assess the magnitude of this, we utilize a set of simulations using v2.0 of the \texttt{Pandeia} \citep{10.1117/12.2231768} simulation engine.  We simulate extended sources with $n=1$ S\'ersic indices across a grid of half-light radii, centered on various positions across an MSA shutter \cite[see also][]{2019MNRAS.486.3290M}.  We then calculate the differential slit losses at the positions of \oiii\ and \hb\ as well \nii\ and \ha\ (and \oiii\ and \oii, for reference) compared to the pipeline correction factor, which is calculated assuming sources are uniform across the open shutter area.

\begin{figure}
    \centering
    \includegraphics[width=0.45\textwidth]{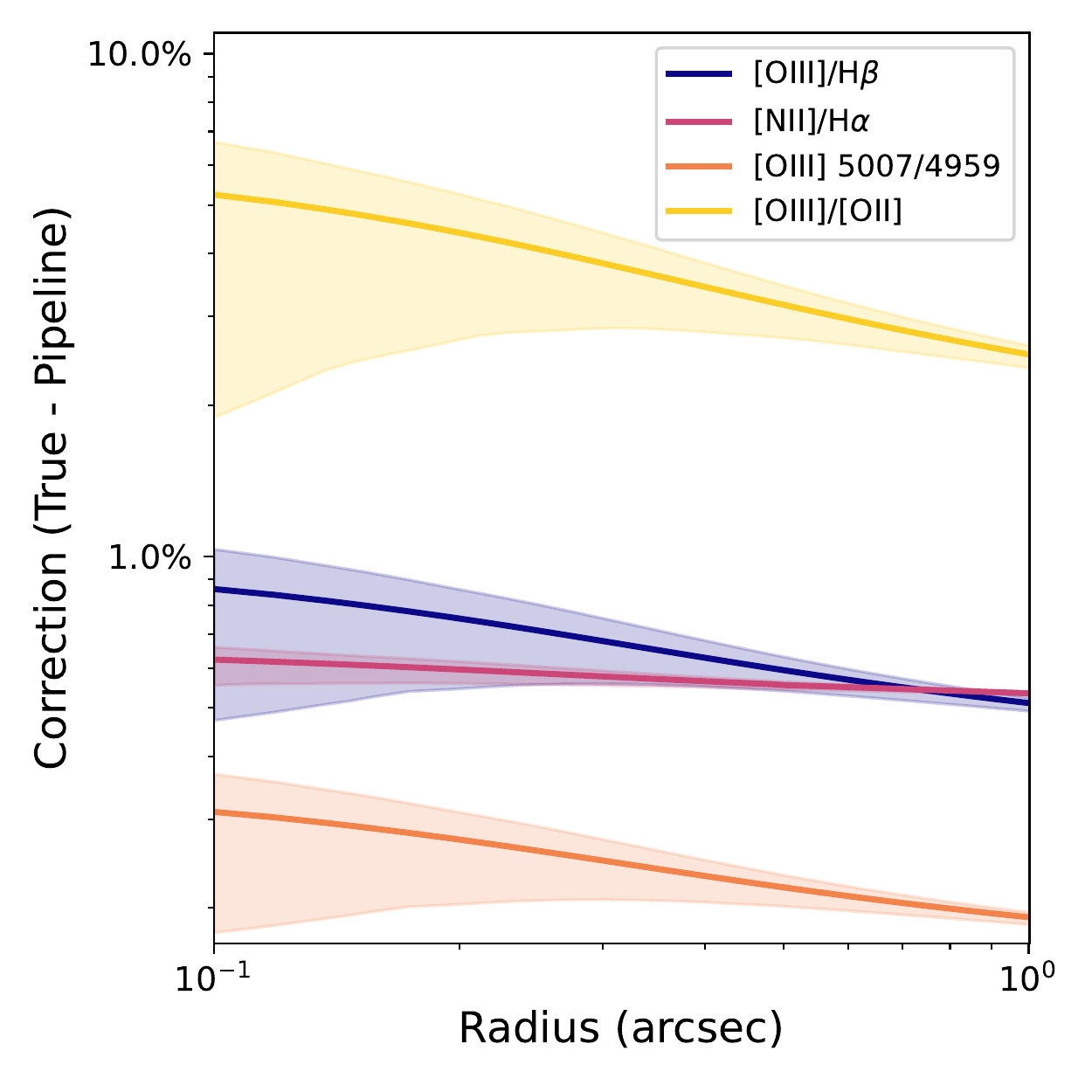} %[..]JWST/MUSE/JWST/analysis/pandeia_ratios_1D.py
    \caption{Difference between the ``true'' correction factor for differential slit losses (from \texttt{Pandeia} simulations, assuming an $n=1$ S\'ersic profile for a $z=5$ galaxy) and the assumption used by the pipeline that sources uniformly illuminate the open shutter area.  The shaded regions denote the +/- 2-$\sigma$ distribution in these values for each line ratio, which encapsulates the variety of spatial offsets possible for a source in the open shutter area.  The NIRSpec pipeline, by default, systematically lowers the observed line ratios as there are larger flux losses at longer wavelengths.  However, the magnitude of this systematic offset cannot explain the offsets we observe between our two stacks in Figure \ref{fig:lineratios}.}
    \label{fig:slitloss}
\end{figure}

The result is shown in Figure \ref{fig:slitloss} as a function of half-light radius, assuming a fiducial redshift of 5.  As an example, at this redshift the pipeline introduces a 0.4\% correction to the \oiii/\hb\ ratio for the slightly larger losses at the wavelength of \oiii.  We calculate that the true correction for sources with an $n=1$ S\'ersic light profile (i.e. not uniform) and a half-light radius of 0$\farcs$1 is 0.9$-$1.5\%, depending on the size and centroid of the source within the open area of an MSA shutter.  Hence the necessary ``correction'' from the measured line ratio to the true one is $\approx 0.5-1.1\%$ (see the $y$-axis of Figure \ref{fig:slitloss}).  The effect is even smaller for the more closely-spaced \nii/\ha\ ratio.  These systematic offsets are not large enough to cause the discrepancy between our two stacks even if, e.g.\  the high-EW sources are systematically smaller, which would cause an artificial suppression of their R3 and N2 ratios, or if there are systematic spatial offsets between the \lya emission centroid and the rest-optical emission lines \cite[e.g.\ ][M. Maseda et al. in prep.]{2016MNRAS.458..449M,2021MNRAS.504.3662L}.  One additional verification comes from observations of both the $\lambda$5007 and $\lambda$4959 components of \oiii: for sources with detections of both lines at $>$2-$\sigma$ the average ratio is 2.96$\pm$0.0937, comparable to the value of 2.98 expected from \citet{2000MNRAS.312..813S}.  The only case where both lines are detected with a signal-to-noise value greater than 2 and the deviation from the expected 2.98 is larger than 2-$\sigma$ is 147002007 (central in Figure \ref{fig:specs}) where there is a negative flux residual from the cosmic ray subtraction near the 4959 emission line in the 2D frame, biasing the boxcar-extracted flux low without properly contributing to an elevated noise level.  We expect improvements to the cosmic ray subtraction in future versions of the pipeline.

A similar systematic effect in our ability to measure line ratios could occur due to the 1D spectral extractions, where the assumption of a fixed extraction window (e.g.\  a boxcar) could result in the preferential loss of flux at long wavelengths compared to short wavelengths.  As we adopt a boxcar width equal to the full open area of the shutter, our 1D extractions are not by themselves introducing a systematic effect to our measurement of the emission line ratios compared to the geometrical effects described above and the nature of the nodded background subtraction which removes flux that extends to large radii.  A more careful ``optimal'' spectral extraction \citep{1986PASP...98..609H} combined with a ``master sky'' background subtraction will be necessary for the precise flux calibration needed for accurate measurement of ratios for distantly-separated emission lines.  

Finally, dust attenuation could have the opposite effect whereby the R3 and N2 ratios would be systematically biased towards higher values.  However, as discussed in \citet{2020MNRAS.493.5120M}, there is little to no dust in galaxies selected based on high \lya EWs.  Moreover, this effect would also be small compared to the observed difference in the two stacks: an $A_V$ of 1 would cause a bias in R3 of only 3$\%$ or 0.01 dex.

\bibliography{ms}

\begin{thebibliography}{}
\expandafter\ifx\csname natexlab\endcsname\relax\def\natexlab#1{#1}\fi
\providecommand{\url}[1]{\href{#1}{#1}}
\providecommand{\dodoi}[1]{doi:~\href{http://doi.org/#1}{\nolinkurl{#1}}}
\providecommand{\doeprint}[1]{\href{http://ascl.net/#1}{\nolinkurl{http://ascl.net/#1}}}
\providecommand{\doarXiv}[1]{\href{https://arxiv.org/abs/#1}{\nolinkurl{https://arxiv.org/abs/#1}}}

\bibitem[{{Astropy Collaboration} {et~al.}(2022){Astropy Collaboration},
  {Price-Whelan}, {Lim}, {Earl}, {Starkman}, {Bradley}, {Shupe}, {Patil},
  {Corrales}, {Brasseur}, {N{\"o}the}, {Donath}, {Tollerud}, {Morris},
  {Ginsburg}, {Vaher}, {Weaver}, {Tocknell}, {Jamieson}, {van Kerkwijk},
  {Robitaille}, {Merry}, {Bachetti}, {G{\"u}nther}, {Aldcroft},
  {Alvarado-Montes}, {Archibald}, {B{\'o}di}, {Bapat}, {Barentsen},
  {Baz{\'a}n}, {Biswas}, {Boquien}, {Burke}, {Cara}, {Cara}, {Conroy},
  {Conseil}, {Craig}, {Cross}, {Cruz}, {D'Eugenio}, {Dencheva}, {Devillepoix},
  {Dietrich}, {Eigenbrot}, {Erben}, {Ferreira}, {Foreman-Mackey}, {Fox},
  {Freij}, {Garg}, {Geda}, {Glattly}, {Gondhalekar}, {Gordon}, {Grant},
  {Greenfield}, {Groener}, {Guest}, {Gurovich}, {Handberg}, {Hart},
  {Hatfield-Dodds}, {Homeier}, {Hosseinzadeh}, {Jenness}, {Jones}, {Joseph},
  {Kalmbach}, {Karamehmetoglu}, {Ka{\l}uszy{\'n}ski}, {Kelley}, {Kern},
  {Kerzendorf}, {Koch}, {Kulumani}, {Lee}, {Ly}, {Ma}, {MacBride}, {Maljaars},
  {Muna}, {Murphy}, {Norman}, {O'Steen}, {Oman}, {Pacifici}, {Pascual},
  {Pascual-Granado}, {Patil}, {Perren}, {Pickering}, {Rastogi}, {Roulston},
  {Ryan}, {Rykoff}, {Sabater}, {Sakurikar}, {Salgado}, {Sanghi}, {Saunders},
  {Savchenko}, {Schwardt}, {Seifert-Eckert}, {Shih}, {Jain}, {Shukla}, {Sick},
  {Simpson}, {Singanamalla}, {Singer}, {Singhal}, {Sinha}, {Sip{\H{o}}cz},
  {Spitler}, {Stansby}, {Streicher}, {{\v{S}}umak}, {Swinbank}, {Taranu},
  {Tewary}, {Tremblay}, {Val-Borro}, {Van Kooten}, {Vasovi{\'c}}, {Verma}, {de
  Miranda Cardoso}, {Williams}, {Wilson}, {Winkel}, {Wood-Vasey}, {Xue},
  {Yoachim}, {Zhang}, {Zonca}, \& {Astropy Project
  Contributors}}]{2022ApJ...935..167A}
{Astropy Collaboration}, {Price-Whelan}, A.~M., {Lim}, P.~L., {et~al.} 2022,
  \apj, 935, 167, \dodoi{10.3847/1538-4357/ac7c74}

\bibitem[{{Bacon} {et~al.}(2010){Bacon}, {Accardo}, {Adjali}, {Anwand},
  {Bauer}, {Biswas}, {Blaizot}, {Boudon}, {Brau-Nogue}, {Brinchmann},
  {Caillier}, {Capoani}, {Carollo}, {Contini}, {Couderc}, {Daguis{\'e}},
  {Deiries}, {Delabre}, {Dreizler}, {Dubois}, {Dupieux}, {Dupuy}, {Emsellem},
  {Fechner}, {Fleischmann}, {Fran{\c c}ois}, {Gallou}, {Gharsa}, {Glindemann},
  {Gojak}, {Guiderdoni}, {Hansali}, {Hahn}, {Jarno}, {Kelz}, {Koehler},
  {Kosmalski}, {Laurent}, {Le Floch}, {Lilly}, {Lizon}, {Loupias}, {Manescau},
  {Monstein}, {Nicklas}, {Olaya}, {Pares}, {Pasquini}, {P{\'e}contal-Rousset},
  {Pell{\'o}}, {Petit}, {Popow}, {Reiss}, {Remillieux}, {Renault}, {Roth},
  {Rupprecht}, {Serre}, {Schaye}, {Soucail}, {Steinmetz}, {Streicher}, {Stuik},
  {Valentin}, {Vernet}, {Weilbacher}, {Wisotzki}, \&
  {Yerle}}]{2010SPIE.7735E..08B}
{Bacon}, R., {Accardo}, M., {Adjali}, L., {et~al.} 2010, in \procspie, Vol.
  7735, Ground-based and Airborne Instrumentation for Astronomy III, 773508,
  \dodoi{10.1117/12.856027}

\bibitem[{{Bacon} {et~al.}(2017){Bacon}, {Conseil}, {Mary}, {Brinchmann},
  {Shepherd}, {Akhlaghi}, {Weilbacher}, {Piqueras}, {Wisotzki}, {Lagattuta},
  {Epinat}, {Guerou}, {Inami}, {Cantalupo}, {Courbot}, {Contini}, {Richard},
  {Maseda}, {Bouwens}, {Bouch{\'e}}, {Kollatschny}, {Schaye}, {Marino},
  {Pello}, {Herenz}, {Guiderdoni}, \& {Carollo}}]{Paper1}
{Bacon}, R., {Conseil}, S., {Mary}, D., {et~al.} 2017, \aap, 608, A1,
  \dodoi{10.1051/0004-6361/201730833}

\bibitem[{{Bacon} {et~al.}(2023){Bacon}, {Brinchmann}, {Conseil}, {Maseda},
  {Nanayakkara}, {Wendt}, {Bacher}, {Mary}, {Weilbacher}, {Krajnovi{\'c}},
  {Boogaard}, {Bouch{\'e}}, {Contini}, {Epinat}, {Feltre}, {Guo}, {Herenz},
  {Kollatschny}, {Kusakabe}, {Leclercq}, {Michel-Dansac}, {Pello}, {Richard},
  {Roth}, {Salvignol}, {Schaye}, {Steinmetz}, {Tresse}, {Urrutia}, {Verhamme},
  {Vitte}, {Wisotzki}, \& {Zoutendijk}}]{2023AA...670A...4B}
{Bacon}, R., {Brinchmann}, J., {Conseil}, S., {et~al.} 2023, \aap, 670, A4,
  \dodoi{10.1051/0004-6361/202244187}

\bibitem[{{Baldwin} {et~al.}(1981){Baldwin}, {Phillips}, \&
  {Terlevich}}]{1981PASP...93....5B}
{Baldwin}, J.~A., {Phillips}, M.~M., \& {Terlevich}, R. 1981, \pasp, 93, 5,
  \dodoi{10.1086/130766}

\bibitem[{{Berg} {et~al.}(2016){Berg}, {Skillman}, {Henry}, {Erb}, \&
  {Carigi}}]{2016ApJ...827..126B}
{Berg}, D.~A., {Skillman}, E.~D., {Henry}, R. B.~C., {Erb}, D.~K., \& {Carigi},
  L. 2016, \apj, 827, 126, \dodoi{10.3847/0004-637X/827/2/126}

\bibitem[{{Bian} {et~al.}(2018){Bian}, {Kewley}, \&
  {Dopita}}]{2018ApJ...859..175B}
{Bian}, F., {Kewley}, L.~J., \& {Dopita}, M.~A. 2018, \apj, 859, 175,
  \dodoi{10.3847/1538-4357/aabd74}

\bibitem[{{Bouwens} {et~al.}(2014){Bouwens}, {Illingworth}, {Oesch},
  {Labb{\'e}}, {van Dokkum}, {Trenti}, {Franx}, {Smit}, {Gonzalez}, \&
  {Magee}}]{2014ApJ...793..115B}
{Bouwens}, R.~J., {Illingworth}, G.~D., {Oesch}, P.~A., {et~al.} 2014, \apj,
  793, 115, \dodoi{10.1088/0004-637X/793/2/115}

\bibitem[{Brammer(2023)}]{brammer_gabriel_2023_7712834}
Brammer, G. 2023, grizli, 1.8.2,  Zenodo, \dodoi{10.5281/zenodo.7712834}

\bibitem[{{Brinchmann}(2023)}]{2022arXiv220807467B}
{Brinchmann}, J. 2023, \mnras, \dodoi{10.1093/mnras/stad1704}

\bibitem[{Bushouse {et~al.}(2022)Bushouse, Eisenhamer, Dencheva, Davies,
  Greenfield, Morrison, Hodge, Simon, Grumm, Droettboom, Slavich, Sosey, Pauly,
  Miller, Jedrzejewski, Hack, Davis, Crawford, Law, Gordon, Regan, Cara,
  MacDonald, Bradley, Shanahan, Jamieson, Teodoro, \&
  Williams}]{bushouse_howard_2022_7229890}
Bushouse, H., Eisenhamer, J., Dencheva, N., {et~al.} 2022, JWST Calibration
  Pipeline, 1.8.2,  Zenodo, \dodoi{10.5281/zenodo.7229890}

\bibitem[{{Cameron} {et~al.}(2023){Cameron}, {Saxena}, {Bunker}, {D'Eugenio},
  {Carniani}, {Maiolino}, {Curtis-Lake}, {Ferruit}, {Jakobsen}, {Arribas},
  {Bonaventura}, {Charlot}, {Chevallard}, {Curti}, {Looser}, {Maseda}, {Rawle},
  {Rodr{\'\i}guez Del Pino}, {Smit}, {{\"U}bler}, {Willott}, {Witstok},
  {Egami}, {Eisenstein}, {Johnson}, {Hainline}, {Rieke}, {Robertson}, {Stark},
  {Tacchella}, {Williams}, {Bhatawdekar}, {Bowler}, {Boyett}, {Circosta},
  {Helton}, {Jones}, {Kumari}, {Ji}, {Nelson}, {Parlanti}, {Sandles},
  {Scholtz}, \& {Sun}}]{2023arXiv230204298C}
{Cameron}, A.~J., {Saxena}, A., {Bunker}, A.~J., {et~al.} 2023, arXiv e-prints,
  arXiv:2302.04298, \dodoi{10.48550/arXiv.2302.04298}

\bibitem[{{Charlot} \& {Fall}(1991)}]{1991ApJ...378..471C}
{Charlot}, S., \& {Fall}, S.~M. 1991, \apj, 378, 471, \dodoi{10.1086/170448}

\bibitem[{{Cullen} {et~al.}(2020){Cullen}, {McLure}, {Dunlop}, {Carnall},
  {McLeod}, {Shapley}, {Amor{\'\i}n}, {Bolzonella}, {Castellano}, {Cimatti},
  {Cirasuolo}, {Cucciati}, {Fontana}, {Fontanot}, {Garilli}, {Guaita},
  {Jarvis}, {Pentericci}, {Pozzetti}, {Talia}, {Zamorani}, {Calabr{\`o}},
  {Cresci}, {Fynbo}, {Hathi}, {Giavalisco}, {Koekemoer}, {Mannucci}, \&
  {Saxena}}]{2020MNRAS.495.1501C}
{Cullen}, F., {McLure}, R.~J., {Dunlop}, J.~S., {et~al.} 2020, \mnras, 495,
  1501, \dodoi{10.1093/mnras/staa1260}

\bibitem[{{Curti} {et~al.}(2023){Curti}, {D'Eugenio}, {Carniani}, {Maiolino},
  {Sandles}, {Witstok}, {Baker}, {Bennett}, {Piotrowska}, {Tacchella},
  {Charlot}, {Nakajima}, {Maheson}, {Mannucci}, {Amiri}, {Arribas}, {Belfiore},
  {Bonaventura}, {Bunker}, {Chevallard}, {Cresci}, {Curtis-Lake},
  {Hayden-Pawson}, {Jones}, {Kumari}, {Laseter}, {Looser}, {Marconi}, {Maseda},
  {Scholtz}, {Smit}, {{\"U}bler}, \& {Wallace}}]{2023MNRAS.518..425C}
{Curti}, M., {D'Eugenio}, F., {Carniani}, S., {et~al.} 2023, \mnras, 518, 425,
  \dodoi{10.1093/mnras/stac2737}

\bibitem[{{da Cunha} {et~al.}(2008){da Cunha}, {Charlot}, \&
  {Elbaz}}]{2008MNRAS.388.1595D}
{da Cunha}, E., {Charlot}, S., \& {Elbaz}, D. 2008, \mnras, 388, 1595,
  \dodoi{10.1111/j.1365-2966.2008.13535.x}

\bibitem[{{Du} {et~al.}(2021){Du}, {Shapley}, {Topping}, {Reddy}, {Sanders},
  {Coil}, {Kriek}, {Mobasher}, \& {Siana}}]{2021ApJ...920...95D}
{Du}, X., {Shapley}, A.~E., {Topping}, M.~W., {et~al.} 2021, \apj, 920, 95,
  \dodoi{10.3847/1538-4357/ac1273}

\bibitem[{{Erb} {et~al.}(2016){Erb}, {Pettini}, {Steidel}, {Strom}, {Rudie},
  {Trainor}, {Shapley}, \& {Reddy}}]{2016ApJ...830...52E}
{Erb}, D.~K., {Pettini}, M., {Steidel}, C.~C., {et~al.} 2016, \apj, 830, 52,
  \dodoi{10.3847/0004-637X/830/1/52}

\bibitem[{{Fujimoto} {et~al.}(2023){Fujimoto}, {Arrabal Haro}, {Dickinson},
  {Finkelstein}, {Kartaltepe}, {Larson}, {Burgarella}, {Bagley}, {Behroozi},
  {Chworowsky}, {Hirschmann}, {Trump}, {Wilkins}, {Yung}, {Koekemoer},
  {Papovich}, {Pirzkal}, {Ferguson}, {Fontana}, {Grogin}, {Grazian}, {Kewley},
  {Kocevski}, {Lotz}, {Pentericci}, {Ravindranath}, {Somerville}, {Amorin},
  {Backhaus}, {Calabro}, {Casey}, {Cooper}, {Franco}, {Giavalisco}, {Hathi},
  {Harish}, {Hutchison}, {Iyer}, {Jung}, {Lucas}, \&
  {Zavala}}]{2023arXiv230109482F}
{Fujimoto}, S., {Arrabal Haro}, P., {Dickinson}, M., {et~al.} 2023, arXiv
  e-prints, arXiv:2301.09482, \dodoi{10.48550/arXiv.2301.09482}

\bibitem[{{Gutkin} {et~al.}(2016){Gutkin}, {Charlot}, \&
  {Bruzual}}]{2016MNRAS.462.1757G}
{Gutkin}, J., {Charlot}, S., \& {Bruzual}, G. 2016, \mnras, 462, 1757,
  \dodoi{10.1093/mnras/stw1716}

\bibitem[{{Harikane} {et~al.}(2018){Harikane}, {Ouchi}, {Shibuya}, {Kojima},
  {Zhang}, {Itoh}, {Ono}, {Higuchi}, {Inoue}, {Chevallard}, {Capak}, {Nagao},
  {Onodera}, {Faisst}, {Martin}, {Rauch}, {Bruzual}, {Charlot}, {Davidzon},
  {Fujimoto}, {Hilmi}, {Ilbert}, {Lee}, {Matsuoka}, {Silverman}, \&
  {Toft}}]{2018ApJ...859...84H}
{Harikane}, Y., {Ouchi}, M., {Shibuya}, T., {et~al.} 2018, \apj, 859, 84,
  \dodoi{10.3847/1538-4357/aabd80}

\bibitem[{Harris {et~al.}(2020)Harris, Millman, van~der Walt, Gommers,
  Virtanen, Cournapeau, Wieser, Taylor, Berg, Smith, Kern, Picus, Hoyer, van
  Kerkwijk, Brett, Haldane, del R{\'{i}}o, Wiebe, Peterson,
  G{\'{e}}rard-Marchant, Sheppard, Reddy, Weckesser, Abbasi, Gohlke, \&
  Oliphant}]{harris2020array}
Harris, C.~R., Millman, K.~J., van~der Walt, S.~J., {et~al.} 2020, Nature, 585,
  357, \dodoi{10.1038/s41586-020-2649-2}

\bibitem[{{Hashimoto} {et~al.}(2017){Hashimoto}, {Garel}, {Guiderdoni},
  {Drake}, {Bacon}, {Blaizot}, {Richard}, {Leclercq}, {Inami}, {Verhamme},
  {Bouwens}, {Brinchmann}, {Cantalupo}, {Carollo}, {Caruana}, {Herenz},
  {Kerutt}, {Marino}, {Mitchell}, \& {Schaye}}]{2017AA...608A..10H}
{Hashimoto}, T., {Garel}, T., {Guiderdoni}, B., {et~al.} 2017, \aap, 608, A10,
  \dodoi{10.1051/0004-6361/201731579}

\bibitem[{{Herenz} {et~al.}(2017){Herenz}, {Urrutia}, {Wisotzki}, {Kerutt},
  {Saust}, {Werhahn}, {Schmidt}, {Caruana}, {Diener}, {Bacon}, {Brinchmann},
  {Schaye}, {Maseda}, \& {Weilbacher}}]{2017arXiv170508215H}
{Herenz}, E.~C., {Urrutia}, T., {Wisotzki}, L., {et~al.} 2017, ArXiv e-prints.
\newblock \doarXiv{1705.08215}

\bibitem[{{Horne}(1986)}]{1986PASP...98..609H}
{Horne}, K. 1986, \pasp, 98, 609, \dodoi{10.1086/131801}

\bibitem[{Hunter(2007)}]{Hunter:2007}
Hunter, J.~D. 2007, Computing in Science \& Engineering, 9, 90,
  \dodoi{10.1109/MCSE.2007.55}

\bibitem[{{Illingworth} {et~al.}(2013){Illingworth}, {Magee}, {Oesch},
  {Bouwens}, {Labb{\'e}}, {Stiavelli}, {van Dokkum}, {Franx}, {Trenti},
  {Carollo}, \& {Gonzalez}}]{2013ApJS..209....6I}
{Illingworth}, G.~D., {Magee}, D., {Oesch}, P.~A., {et~al.} 2013, \apjs, 209,
  6, \dodoi{10.1088/0067-0049/209/1/6}

\bibitem[{{Inami} {et~al.}(2017){Inami}, {Bacon}, {Brinchmann}, {Richard},
  {Contini}, \& {Conseil}}]{Paper2}
{Inami}, H., {Bacon}, R., {Brinchmann}, J., {et~al.} 2017, A\&A, Submitted
  (MUSE UDF SI)

\bibitem[{{Inoue}(2011)}]{2011MNRAS.415.2920I}
{Inoue}, A.~K. 2011, \mnras, 415, 2920,
  \dodoi{10.1111/j.1365-2966.2011.18906.x}

\bibitem[{{Karakla} {et~al.}(2014){Karakla}, {Shyrokov}, {Pontoppidan}, {Beck},
  {Gilbert}, {Valenti}, {Kassin}, \& {Soderblom}}]{2014SPIE.9149E..1ZK}
{Karakla}, D., {Shyrokov}, A., {Pontoppidan}, K., {et~al.} 2014, in Society of
  Photo-Optical Instrumentation Engineers (SPIE) Conference Series, Vol. 9149,
  Observatory Operations: Strategies, Processes, and Systems V, ed. A.~B.
  {Peck}, C.~R. {Benn}, \& R.~L. {Seaman}, 91491Z, \dodoi{10.1117/12.2056387}

\bibitem[{{Kerutt} {et~al.}(2022){Kerutt}, {Wisotzki}, {Verhamme}, {Schmidt},
  {Leclercq}, {Herenz}, {Urrutia}, {Garel}, {Hashimoto}, {Maseda}, {Matthee},
  {Kusakabe}, {Schaye}, {Richard}, {Guiderdoni}, {Mauerhofer}, {Nanayakkara},
  \& {Vitte}}]{2022AA...659A.183K}
{Kerutt}, J., {Wisotzki}, L., {Verhamme}, A., {et~al.} 2022, \aap, 659, A183,
  \dodoi{10.1051/0004-6361/202141900}

\bibitem[{{Kewley} {et~al.}(2013){Kewley}, {Dopita}, {Leitherer}, {Dav{\'e}},
  {Yuan}, {Allen}, {Groves}, \& {Sutherland}}]{2013ApJ...774..100K}
{Kewley}, L.~J., {Dopita}, M.~A., {Leitherer}, C., {et~al.} 2013, \apj, 774,
  100, \dodoi{10.1088/0004-637X/774/2/100}

\bibitem[{{Kewley} {et~al.}(2001){Kewley}, {Dopita}, {Sutherland}, {Heisler},
  \& {Trevena}}]{2001ApJ...556..121K}
{Kewley}, L.~J., {Dopita}, M.~A., {Sutherland}, R.~S., {Heisler}, C.~A., \&
  {Trevena}, J. 2001, \apj, 556, 121, \dodoi{10.1086/321545}

\bibitem[{{Laseter} {et~al.}(2022){Laseter}, {Barger}, {Cowie}, \&
  {Taylor}}]{2022ApJ...935..150L}
{Laseter}, I.~H., {Barger}, A.~J., {Cowie}, L.~L., \& {Taylor}, A.~J. 2022,
  \apj, 935, 150, \dodoi{10.3847/1538-4357/ac81c7}

\bibitem[{{Laseter} {et~al.}(2023){Laseter}, {Maseda}, {Curti}, {Maiolino},
  {D'Eugenio}, {Cameron}, {Looser}, {Arribas}, {Baker}, {Bhatawdekar},
  {Boyett}, {Bunker}, {Carniani}, {Charlot}, {Chevallard}, {Curtis-lake},
  {Egami}, {Eisenstein}, {Hainline}, {Hausen}, {Ji}, {Kumari}, {Perna},
  {Rawle}, {Rix}, {Robertson}, {Rodr{\'\i}guez Del Pino}, {Sandles}, {Scholtz},
  {Smit}, {Tacchella}, {{\"U}bler}, {Williams}, {Willott}, \&
  {Witstok}}]{2023arXiv230603120L}
{Laseter}, I.~H., {Maseda}, M.~V., {Curti}, M., {et~al.} 2023, arXiv e-prints,
  arXiv:2306.03120, \dodoi{10.48550/arXiv.2306.03120}

\bibitem[{{Lemaux} {et~al.}(2021){Lemaux}, {Fuller}, {Brada{\v{c}}},
  {Pentericci}, {Hoag}, {Strait}, {Treu}, {Alvarez}, {Bolan}, {Gandhi},
  {Huang}, {Jones}, {Mason}, {Pelliccia}, {Ribeiro}, {Ryan}, {Schmidt},
  {Vanzella}, {Khusanova}, {Le F{\`e}vre}, {Guaita}, {Hathi}, {Koekemoer}, \&
  {Pforr}}]{2021MNRAS.504.3662L}
{Lemaux}, B.~C., {Fuller}, S., {Brada{\v{c}}}, M., {et~al.} 2021, \mnras, 504,
  3662, \dodoi{10.1093/mnras/stab924}

\bibitem[{{Malhotra} \& {Rhoads}(2002)}]{2002ApJ...565L..71M}
{Malhotra}, S., \& {Rhoads}, J.~E. 2002, \apjl, 565, L71,
  \dodoi{10.1086/338980}

\bibitem[{{Maseda} {et~al.}(2019){Maseda}, {Franx}, {Chevallard}, \&
  {Curtis-Lake}}]{2019MNRAS.486.3290M}
{Maseda}, M.~V., {Franx}, M., {Chevallard}, J., \& {Curtis-Lake}, E. 2019,
  \mnras, 486, 3290, \dodoi{10.1093/mnras/stz818}

\bibitem[{{Maseda} {et~al.}(2018){Maseda}, {Bacon}, {Franx}, {Brinchmann},
  {Schaye}, {Boogaard}, {Bouch{\'e}}, {Bouwens}, {Cantalupo}, {Contini},
  {Hashimoto}, {Inami}, {Marino}, {Muzahid}, {Nanayakkara}, {Richard},
  {Schmidt}, {Verhamme}, \& {Wisotzki}}]{2018ApJ...865L...1M}
{Maseda}, M.~V., {Bacon}, R., {Franx}, M., {et~al.} 2018, \apjl, 865, L1,
  \dodoi{10.3847/2041-8213/aade4b}

\bibitem[{{Maseda} {et~al.}(2020){Maseda}, {Bacon}, {Lam}, {Matthee},
  {Brinchmann}, {Schaye}, {Labbe}, {Schmidt}, {Boogaard}, {Bouwens},
  {Cantalupo}, {Franx}, {Hashimoto}, {Inami}, {Kusakabe}, {Mahler},
  {Nanayakkara}, {Richard}, \& {Wisotzki}}]{2020MNRAS.493.5120M}
{Maseda}, M.~V., {Bacon}, R., {Lam}, D., {et~al.} 2020, \mnras, 493, 5120,
  \dodoi{10.1093/mnras/staa622}

\bibitem[{{Matthee} {et~al.}(2022){Matthee}, {Mackenzie}, {Simcoe}, {Kashino},
  {Lilly}, {Bordoloi}, \& {Eilers}}]{2022arXiv221108255M}
{Matthee}, J., {Mackenzie}, R., {Simcoe}, R.~A., {et~al.} 2022, arXiv e-prints,
  arXiv:2211.08255, \dodoi{10.48550/arXiv.2211.08255}

\bibitem[{{Matthee} {et~al.}(2016){Matthee}, {Sobral}, {Oteo}, {Best}, {Smail},
  {R{\"o}ttgering}, \& {Paulino-Afonso}}]{2016MNRAS.458..449M}
{Matthee}, J., {Sobral}, D., {Oteo}, I., {et~al.} 2016, \mnras, 458, 449,
  \dodoi{10.1093/mnras/stw322}

\bibitem[{{Matthee} {et~al.}(2021){Matthee}, {Sobral}, {Hayes}, {Pezzulli},
  {Gronke}, {Schaerer}, {Naidu}, {R{\"o}ttgering}, {Calhau}, {Paulino-Afonso},
  {Santos}, \& {Amor{\'\i}n}}]{2021MNRAS.505.1382M}
{Matthee}, J., {Sobral}, D., {Hayes}, M., {et~al.} 2021, \mnras, 505, 1382,
  \dodoi{10.1093/mnras/stab1304}

\bibitem[{{Momcheva} {et~al.}(2016){Momcheva}, {Brammer}, {van Dokkum},
  {Skelton}, {Whitaker}, {Nelson}, {Fumagalli}, {Maseda}, {Leja}, {Franx},
  {Rix}, {Bezanson}, {Da Cunha}, {Dickey}, {F{\"o}rster Schreiber},
  {Illingworth}, {Kriek}, {Labb{\'e}}, {Ulf Lange}, {Lundgren}, {Magee},
  {Marchesini}, {Oesch}, {Pacifici}, {Patel}, {Price}, {Tal}, {Wake}, {van der
  Wel}, \& {Wuyts}}]{2016ApJS..225...27M}
{Momcheva}, I.~G., {Brammer}, G.~B., {van Dokkum}, P.~G., {et~al.} 2016, \apjs,
  225, 27, \dodoi{10.3847/0067-0049/225/2/27}

\bibitem[{{Muzahid} {et~al.}(2020){Muzahid}, {Schaye}, {Marino}, {Cantalupo},
  {Brinchmann}, {Contini}, {Wendt}, {Wisotzki}, {Zabl}, {Bouch{\'e}},
  {Akhlaghi}, {Chen}, {Claeyssens}, {Johnson}, {Leclercq}, {Maseda}, {Matthee},
  {Richard}, {Urrutia}, \& {Verhamme}}]{2020MNRAS.496.1013M}
{Muzahid}, S., {Schaye}, J., {Marino}, R.~A., {et~al.} 2020, \mnras, 496, 1013,
  \dodoi{10.1093/mnras/staa1347}

\bibitem[{{Nakajima} {et~al.}(2023){Nakajima}, {Ouchi}, {Isobe}, {Harikane},
  {Zhang}, {Ono}, {Umeda}, \& {Oguri}}]{2023arXiv230112825N}
{Nakajima}, K., {Ouchi}, M., {Isobe}, Y., {et~al.} 2023, arXiv e-prints,
  arXiv:2301.12825, \dodoi{10.48550/arXiv.2301.12825}

\bibitem[{{Partridge} \& {Peebles}(1967)}]{1967ApJ...148..377P}
{Partridge}, R.~B., \& {Peebles}, P.~J.~E. 1967, \apj, 148, 377,
  \dodoi{10.1086/149161}

\bibitem[{{Pontoppidan} {et~al.}(2016){Pontoppidan}, {Pickering}, {Laidler},
  {Gilbert}, {Sontag}, {Slocum}, {Sienkiewicz}, {Hanley}, {Earl}, {Pueyo},
  {Ravindranath}, {Karakla}, {Robberto}, {Noriega-Crespo}, \&
  {Barker}}]{10.1117/12.2231768}
{Pontoppidan}, K.~M., {Pickering}, T.~E., {Laidler}, V.~G., {et~al.} 2016,
  Pandeia: a multi-mission exposure time calculator for JWST and WFIRST,
  \dodoi{10.1117/12.2231768}

\bibitem[{{Prochaska} {et~al.}(2020){Prochaska}, {Hennawi}, {Westfall},
  {Cooke}, {Wang}, {Hsyu}, {Davies}, \& {Farina}}]{pypeit:joss_arXiv}
{Prochaska}, J.~X., {Hennawi}, J.~F., {Westfall}, K.~B., {et~al.} 2020, arXiv
  e-prints, arXiv:2005.06505.
\newblock \doarXiv{2005.06505}

\bibitem[{Prochaska {et~al.}(2020)Prochaska, Hennawi, Westfall, Cooke, Wang,
  Hsyu, Davies, Farina, \& Pelliccia}]{pypeit:joss_pub}
Prochaska, J.~X., Hennawi, J.~F., Westfall, K.~B., {et~al.} 2020, Journal of
  Open Source Software, 5, 2308, \dodoi{10.21105/joss.02308}

\bibitem[{{Prochaska} {et~al.}(2020){Prochaska}, {Hennawi}, {Cooke},
  {Westfall}, {Wang}, {EmAstro}, {Tiffanyhsyu}, {Wasserman}, {Villaume},
  {Marijana777}, {Schindler}, {Young}, {Simha}, {Wilde}, {Tejos}, {Isbell},
  {Fl{\"o}rs}, {Sandford}, {Vasovi{\'c}}, {Betts}, \& {Holden}}]{pypeit:zenodo}
{Prochaska}, J.~X., {Hennawi}, J., {Cooke}, R., {et~al.} 2020, {pypeit/PypeIt:
  Release 1.0.0}, v1.0.0,  Zenodo, \dodoi{10.5281/zenodo.3743493}

\bibitem[{{Rafelski} {et~al.}(2015){Rafelski}, {Teplitz}, {Gardner}, {Coe},
  {Bond}, {Koekemoer}, {Grogin}, {Kurczynski}, {McGrath}, {Bourque}, {Atek},
  {Brown}, {Colbert}, {Codoreanu}, {Ferguson}, {Finkelstein}, {Gawiser},
  {Giavalisco}, {Gronwall}, {Hanish}, {Lee}, {Mehta}, {de Mello},
  {Ravindranath}, {Ryan}, {Scarlata}, {Siana}, {Soto}, \&
  {Voyer}}]{2015AJ....150...31R}
{Rafelski}, M., {Teplitz}, H.~I., {Gardner}, J.~P., {et~al.} 2015, \aj, 150,
  31, \dodoi{10.1088/0004-6256/150/1/31}

\bibitem[{{Raiter} {et~al.}(2010){Raiter}, {Schaerer}, \&
  {Fosbury}}]{2010AA...523A..64R}
{Raiter}, A., {Schaerer}, D., \& {Fosbury}, R.~A.~E. 2010, \aap, 523, A64,
  \dodoi{10.1051/0004-6361/201015236}

\bibitem[{{Rawle} {et~al.}(2022){Rawle}, {Giardino}, {Franz}, {Rapp}, {te
  Plate}, {Zincke}, {Abul-Huda}, {Alves de Oliveira}, {Bechtold}, {Beck},
  {Birkmann}, {B{\"o}ker}, {Ehrenwinkler}, {Ferruit}, {Garland}, {Jakobsen},
  {Karakla}, {Karl}, {Keyes}, {Koehler}, {Nimisha}, {L{\"u}tzgendorf},
  {Manjavacas}, {Marston}, {Moseley}, {Mosner}, {Muzerolle}, {Ogle},
  {Proffitt}, {Sabbi}, {Sirianni}, {Wahlgren}, {Wislowski}, {Wright}, {Wu}, \&
  {Zeidler}}]{2022SPIE12180E..3RR}
{Rawle}, T.~D., {Giardino}, G., {Franz}, D.~E., {et~al.} 2022, in Society of
  Photo-Optical Instrumentation Engineers (SPIE) Conference Series, Vol. 12180,
  Space Telescopes and Instrumentation 2022: Optical, Infrared, and Millimeter
  Wave, ed. L.~E. {Coyle}, S.~{Matsuura}, \& M.~D. {Perrin}, 121803R,
  \dodoi{10.1117/12.2629231}

\bibitem[{{Roberts-Borsani} {et~al.}(2022){Roberts-Borsani}, {Treu}, {Chen},
  {Morishita}, {Vanzella}, {Zitrin}, {Bergamini}, {Castellano}, {Fontana},
  {Grillo}, {Kelly}, {Merlin}, {Paris}, {Rosati}, {Acebron}, {Bonchi},
  {Boyett}, {Bradac}, {Broadhurst}, {Calabro}, {Diego}, {Dressler}, {Furtak},
  {Filippenko}, {Glazebrook}, {Koekemoer}, {Leethochawalit}, {Malkan}, {Mason},
  {Mercurio}, {Metha}, {Nanayakkara}, {Pentericci}, {Pierel}, {Rieck}, {Roy},
  {Santini}, {Strait}, {Strausbaugh}, {Trenti}, {Vulcani}, {Wang}, {Wang},
  {Windhorst}, \& {Yang}}]{2022arXiv221015639R}
{Roberts-Borsani}, G., {Treu}, T., {Chen}, W., {et~al.} 2022, arXiv e-prints,
  arXiv:2210.15639, \dodoi{10.48550/arXiv.2210.15639}

\bibitem[{{Sanders} {et~al.}(2023{\natexlab{a}}){Sanders}, {Shapley},
  {Topping}, {Reddy}, \& {Brammer}}]{2023arXiv230106696S}
{Sanders}, R.~L., {Shapley}, A.~E., {Topping}, M.~W., {Reddy}, N.~A., \&
  {Brammer}, G.~B. 2023{\natexlab{a}}, arXiv e-prints, arXiv:2301.06696,
  \dodoi{10.48550/arXiv.2301.06696}

\bibitem[{{Sanders} {et~al.}(2023{\natexlab{b}}){Sanders}, {Shapley},
  {Topping}, {Reddy}, \& {Brammer}}]{2023arXiv230308149S}
---. 2023{\natexlab{b}}, arXiv e-prints, arXiv:2303.08149,
  \dodoi{10.48550/arXiv.2303.08149}

\bibitem[{{Schaerer}(2003)}]{2003AA...397..527S}
{Schaerer}, D. 2003, \aap, 397, 527, \dodoi{10.1051/0004-6361:20021525}

\bibitem[{{Schaerer} {et~al.}(2022){Schaerer}, {Marques-Chaves}, {Barrufet},
  {Oesch}, {Izotov}, {Naidu}, {Guseva}, \& {Brammer}}]{2022AA...665L...4S}
{Schaerer}, D., {Marques-Chaves}, R., {Barrufet}, L., {et~al.} 2022, \aap, 665,
  L4, \dodoi{10.1051/0004-6361/202244556}

\bibitem[{{Shapley} {et~al.}(2003){Shapley}, {Steidel}, {Pettini}, \&
  {Adelberger}}]{2003ApJ...588...65S}
{Shapley}, A.~E., {Steidel}, C.~C., {Pettini}, M., \& {Adelberger}, K.~L. 2003,
  \apj, 588, 65, \dodoi{10.1086/373922}

\bibitem[{{Simmonds} {et~al.}(2023){Simmonds}, {Tacchella}, {Maseda},
  {Williams}, {Baker}, {Witten}, {Johnson}, {Robertson}, {Saxena}, {Sun},
  {Witstok}, {Bhatawdekar}, {Boyett}, {Bunker}, {Charlot}, {Curtis-Lake},
  {Egami}, {Eisenstein}, {Ji}, {Maiolino}, {Sandles}, {Smit}, {{\"U}bler}, \&
  {Willott}}]{2023arXiv230307931S}
{Simmonds}, C., {Tacchella}, S., {Maseda}, M.~V., {et~al.} 2023, arXiv
  e-prints, arXiv:2303.07931, \dodoi{10.48550/arXiv.2303.07931}

\bibitem[{{Skelton} {et~al.}(2014){Skelton}, {Whitaker}, {Momcheva}, {Brammer},
  {van Dokkum}, {Labb{\'e}}, {Franx}, {van der Wel}, {Bezanson}, {Da Cunha},
  {Fumagalli}, {F{\"o}rster Schreiber}, {Kriek}, {Leja}, {Lundgren}, {Magee},
  {Marchesini}, {Maseda}, {Nelson}, {Oesch}, {Pacifici}, {Patel}, {Price},
  {Rix}, {Tal}, {Wake}, \& {Wuyts}}]{2014ApJS..214...24S}
{Skelton}, R.~E., {Whitaker}, K.~E., {Momcheva}, I.~G., {et~al.} 2014, \apjs,
  214, 24, \dodoi{10.1088/0067-0049/214/2/24}

\bibitem[{{Storey} \& {Zeippen}(2000)}]{2000MNRAS.312..813S}
{Storey}, P.~J., \& {Zeippen}, C.~J. 2000, \mnras, 312, 813,
  \dodoi{10.1046/j.1365-8711.2000.03184.x}

\bibitem[{{Strom} {et~al.}(2017){Strom}, {Steidel}, {Rudie}, {Trainor},
  {Pettini}, \& {Reddy}}]{2017ApJ...836..164S}
{Strom}, A.~L., {Steidel}, C.~C., {Rudie}, G.~C., {et~al.} 2017, \apj, 836,
  164, \dodoi{10.3847/1538-4357/836/2/164}

\bibitem[{{Taylor} {et~al.}(2022){Taylor}, {Barger}, \&
  {Cowie}}]{2022ApJ...939L...3T}
{Taylor}, A.~J., {Barger}, A.~J., \& {Cowie}, L.~L. 2022, \apjl, 939, L3,
  \dodoi{10.3847/2041-8213/ac959d}

\bibitem[{{Trainor} {et~al.}(2015){Trainor}, {Steidel}, {Strom}, \&
  {Rudie}}]{2015ApJ...809...89T}
{Trainor}, R.~F., {Steidel}, C.~C., {Strom}, A.~L., \& {Rudie}, G.~C. 2015,
  \apj, 809, 89, \dodoi{10.1088/0004-637X/809/1/89}

\bibitem[{{Trainor} {et~al.}(2016){Trainor}, {Strom}, {Steidel}, \&
  {Rudie}}]{2016ApJ...832..171T}
{Trainor}, R.~F., {Strom}, A.~L., {Steidel}, C.~C., \& {Rudie}, G.~C. 2016,
  \apj, 832, 171, \dodoi{10.3847/0004-637X/832/2/171}

\bibitem[{{Trump} {et~al.}(2022){Trump}, {Arrabal Haro}, {Simons}, {Backhaus},
  {Amor{\'\i}n}, {Dickinson}, {Fern{\'a}ndez}, {Papovich}, {Nicholls},
  {Kewley}, {Brunker}, {Salzer}, {Wilkins}, {Almaini}, {Bagley}, {Berg},
  {Bhatawdekar}, {Bisigello}, {Buat}, {Burgarella}, {Calabr{\`o}}, {Casey},
  {Ciesla}, {Cleri}, {Cole}, {Cooper}, {Cooray}, {Costantin}, {Ferguson},
  {Finkelstein}, {Fujimoto}, {Gardner}, {Gawiser}, {Giavalisco}, {Grazian},
  {Grogin}, {Hathi}, {Hirschmann}, {Holwerda}, {Huertas-Company}, {Hutchison},
  {Jogee}, {Juneau}, {Jung}, {Kartaltepe}, {Kirkpatrick}, {Koekemoer}, {Lotz},
  {Lucas}, {Magnelli}, {Matharu}, {P{\'e}rez-Gonz{\'a}lez}, {Pirzkal},
  {Rafelski}, {Rose}, {Seill{\'e}}, {Somerville}, {Straughn}, {Tacchella},
  {Vanderhoof}, {Weiner}, {Wuyts}, {Yung}, \& {Zavala}}]{2022arXiv220712388T}
{Trump}, J.~R., {Arrabal Haro}, P., {Simons}, R.~C., {et~al.} 2022, arXiv
  e-prints, arXiv:2207.12388.
\newblock \doarXiv{2207.12388}

\bibitem[{{Umeda} {et~al.}(2022){Umeda}, {Ouchi}, {Nakajima}, {Isobe},
  {Aoyama}, {Harikane}, {Ono}, \& {Matsumoto}}]{2022ApJ...930...37U}
{Umeda}, H., {Ouchi}, M., {Nakajima}, K., {et~al.} 2022, \apj, 930, 37,
  \dodoi{10.3847/1538-4357/ac602d}

\bibitem[{{Urrutia} {et~al.}(2019){Urrutia}, {Wisotzki}, {Kerutt}, {Schmidt},
  {Herenz}, {Klar}, {Saust}, {Werhahn}, {Diener}, {Caruana}, {Krajnovi{\'c}},
  {Bacon}, {Boogaard}, {Brinchmann}, {Enke}, {Maseda}, {Nanayakkara},
  {Richard}, {Steinmetz}, \& {Weilbacher}}]{2019AA...624A.141U}
{Urrutia}, T., {Wisotzki}, L., {Kerutt}, J., {et~al.} 2019, \aap, 624, A141,
  \dodoi{10.1051/0004-6361/201834656}

\bibitem[{{Veilleux} \& {Osterbrock}(1987)}]{1987ApJS...63..295V}
{Veilleux}, S., \& {Osterbrock}, D.~E. 1987, \apjs, 63, 295,
  \dodoi{10.1086/191166}

\bibitem[{{Wang} {et~al.}(2022){Wang}, {Cheng}, {Ge}, {Meng}, {Daddi}, {Yan},
  {Jones}, {Malkan}, {Arrabal Haro}, {Brammer}, \&
  {Oguri}}]{2022arXiv221204476W}
{Wang}, X., {Cheng}, C., {Ge}, J., {et~al.} 2022, arXiv e-prints,
  arXiv:2212.04476, \dodoi{10.48550/arXiv.2212.04476}

\bibitem[{{Zackrisson} {et~al.}(2011){Zackrisson}, {Inoue}, {Rydberg}, \&
  {Duval}}]{2011MNRAS.418L.104Z}
{Zackrisson}, E., {Inoue}, A.~K., {Rydberg}, C.-E., \& {Duval}, F. 2011,
  \mnras, 418, L104, \dodoi{10.1111/j.1745-3933.2011.01153.x}

\end{thebibliography}

\end{document}